\documentclass[preprint,nofootinbib,aps,superscriptaddress]{revtex4}
\usepackage{color,graphicx,shortvrb,epsfig}
\newcommand{\beq}{\begin{equation}}
\newcommand{\eeq}{\end{equation}}
\newcommand{\beqa}{\begin{eqnarray}}
\newcommand{\eeqa}{\end{eqnarray}}
\newcommand{\no}{\nonumber}
\def\lsim{\mathrel{\rlap{\lower4pt\hbox{\hskip1pt$\sim$}}
    \raise1pt\hbox{$<$}}}         %less than or approx. symbol
\def\gsim{\mathrel{\rlap{\lower4pt\hbox{\hskip1pt$\sim$}}
    \raise1pt\hbox{$>$}}}         %greater than or approx. symbol
\newcommand{\re}[1]{\ensuremath{{\cal R}e(#1)}}
\newcommand{\im}[1]{\ensuremath{{\cal I}m(#1)}}
\newcommand{\Bz}{{B^0}}
\newcommand{\Bzb}{\overline{B}{}^0}

\newcommand{\CP}{CP\ }

\newcommand{\Bbar}{\overline{B}}

\newcommand{\Heff}{{\cal H}}
\newcommand{\Meff}{M}
\newcommand{\Geff}{\Gamma}

\newcommand{\fb}{\overline{f}}
\newcommand{\f}{f}

\begin{document}

%\preprint{\vbox{\hbox{WIS/2/03-Mar-DPP}
%  \hbox{hep-ph/0303171}}}

\vspace*{1cm}

\title{\boldmath Probing new physics with flavor physics\\
  (and probing flavor physics with new physics)\footnote{
Lectures given at PiTP 2007, ``Standard Model and Beyond'', IAS,
  Princeton, USA, July 16--25 2007, and at the 2007 CERN-FERMILAB
  Hadron Collider Physics 
Summer School, CERN, Geneva, Switzerland, June 6--15 2007.}}

\author{Yosef Nir}\email{yosef.nir@weizmann.ac.il}
\affiliation{Department of Particle Physics \\
  Weizmann Institute of Science, Rehovot 76100, Israel}  

%\date{\today}
%\pacs{12.10.Dm, 12.10.Kt, 98.80.Cq}

\begin{abstract}
This is a written version of a series of lectures aimed at graduate
students and postdoctoral fellows in particle theory/string
theory/particle experiment familiar with the basics of the Standard
Model. We begin with an overview of flavor physics
and its implications for new physics. We emphasize the ``new physics
flavor puzzle''. Then, we give four specific examples of flavor
measurements and the lessons that have been (or can be) drawn from
them: (i) Charm physics: lessons for supersymmetry from the upper
bound on $\Delta m_D$. (ii) Bottom physics: model independent lessons
on the KM mechanism and on new physics in $B^0-\overline{B}{}^0$
mixing from $S_{\psi K_S}$. (iii) Top physics and beyond: testing
minimal flavor violation at the LHC. (iv) Neutrino physics:
interpreting the data on neutrino masses and mixing within flavor
models. 
\end{abstract}

\maketitle
%\tableofcontents

%%%%%%%%%%%%%%%%%%%%%%%%%%%
%%%%%%%%%%   I   %%%%%%%%%%
\section{Introduction}
The Standard Model fermions appear in three generations. {\it Flavor
physics} describes interactions that distinguish between the fermion
generations. 

The fermions experience two types of interactions: gauge interactions,
where two fermions couple to a gauge boson, and Yukawa interactions,
where two fermions couple to a scalar.  In the {\it interaction
  basis}, gauge interactions are diagonal and universal, namely
described by a single gauge coupling for each type of interaction
($g_s$, $g$, and $g'$). By definition, there are no gauge couplings
between interaction eigenstates of different generations. The Yukawa
interactions are, however, quite complicated in the interaction basis.
In particular, there are Yukawa couplings that involve fermions of
different generations and, consequently, the interaction eigenstates
do not have well-defined masses. {\it Flavor physics} here refers to
the part of the Standard Model that depends on the Yukawa couplings.

In the {\it mass basis}, Yukawa interactions are diagonal (in the
Standard Model, its single-Higgs extensions and even with extended
Higgs sector subject to natural flavor conservation), but not
universal. The mass eigenstates have, by definition, well-defined
masses. The interactions related to spontaneously broken symmetries
are, however, quite complicated in the mass basis. In particular, the
interactions of the charged weak force carriers $W^\pm$ are not
diagonal, that is, they {\it mix} quarks of different generations. (In
extensions of the Standard Model, with $SU(2)_{\rm L}$-singlet
left-handed quarks, or $SU(2)_{\rm L}$-doublet right-handed quarks,
also the $Z$-couplings involve mixing.) {\it Flavor physics} here
refers to fermion masses and mixings.

Why is flavor physics interesting?

\begin{itemize}
\item Flavor physics can discover new physics or probe it before it is
  directly observed in experiments. Here are some examples from the
  past:
\begin{itemize}
\item The smallness of $\frac{\Gamma(K_L\to\mu^+\mu^-)}
  {\Gamma(K^+\to\mu^+\nu)}$ led to predicting a fourth (the charm)
  quark;
\item The size of $\Delta m_K$ led to a successful prediction of the
  charm mass;
\item The size of $\Delta m_B$ led to a successful prediction of the
  top mass;
\item The measurement of $\varepsilon_K$ led to predicting the third
  generation.
\end{itemize}
\item CP violation is closely related to flavor physics. Within the
  Standard Model, there is a single CP violating parameter, the
  Kobayashi-Maskawa phase $\delta_{\rm KM}$ \cite{Kobayashi:1973fv}.
  Baryogenesis tells us, however, that there must exist new sources of
  CP violation. Measurements of CP violation in flavor changing
  processes might provide evidence for such sources.
\item The fine-tuning problem of the Higgs mass, and the puzzle of the
  dark matter imply that there exists new physics at, or below, the
  TeV scale. If such new physics had a generic flavor structure, it
  would contribute to flavor changing neutral current (FCNC) processes
  orders of magnitude above the observed rates. The question of why
  this does not happen constitutes the {\it new physics flavor
    puzzle}.
\item Most of the charged fermion flavor parameters are small and
  hierarchical. The Standard Model does not provide any explanation of
  these features. This is the {\it Standard Model flavor puzzle}.  The
  puzzle became even deeper after neutrino masses and mixings were
  measured because, so far, neither smallness nor hierarchy in these
  parameters have been established.
\end{itemize}
In these lectures, we discuss four specific measurements that
relate to the four points above:
\begin{itemize}
\item We show how measurements of $D^0-\overline{D}{}^0$ mixing allow
  us to explore supersymmetry and, in particular, give evidence that
  if there are squarks below the TeV scale, they must be
  quasi-degenerate (Section \ref{sec:dmix}).
\item We explain how the measurement of the CP asymmetry in $B\to
  J/\psi K_S$ decays gives evidence that the KM mechanism is the
  dominant source of the observed CP violation, and quantitatively
  constrains the amount of new physics in $B^0-\overline{B}{}^0$
  mixing (Section \ref{sec:bmix}).
\item We present the idea of minimal flavor violation as a solution to
  the new physics flavor problem, and argue that the ATLAS and CMS
  experiments may be able to test this solution (Section \ref{sec:lhc}).
\item We describe the extraction of four neutrino parameters from
  measurements related to atmospheric and solar neutrinos, and explain
  their impact on models that aim to explain the Standard Model flavor
  puzzle (Section \ref{sec:nu}). 
\end{itemize}

%%%%%%%%%%%%%%%%%%%%%%%%%%%
%%%%%%%%%   II   %%%%%%%%%%
\section{Flavor in the Standard Model}
\label{smfor}
A model of elementary particles and their interactions is defined
by the following ingredients: (i) The symmetries of the Lagrangian and
the pattern of spontaneous symmetry breaking; (ii) The representations
of fermions and scalars. The Standard Model (SM) is defined as follows: 
(i) The gauge symmetry is
\beq\label{smsym}
G_{\rm SM}=SU(3)_{\rm C}\times SU(2)_{\rm L}\times U(1)_{\rm Y}.
\eeq
It is spontaneously broken by the VEV of a single Higgs scalar,
$\phi(1,2)_{1/2}$ ($\langle\phi^0\rangle=v/\sqrt{2}$):
\beq\label{smssb}
G_{\rm SM} \to SU(3)_{\rm C}\times U(1)_{\rm EM}.
\eeq
(ii) There are three fermion generations, each consisting of five 
representations of $G_{\rm SM}$:
\beq\label{ferrep}
Q_{Li}(3,2)_{+1/6},\ \ U_{Ri}(3,1)_{+2/3},\ \ 
D_{Ri}(3,1)_{-1/3},\ \ L_{Li}(1,2)_{-1/2},\ \ E_{Ri}(1,1)_{-1}.
\eeq
%

%%%%%%%%%%%%%
\subsection{The interactions basis}
The Standard Model Lagrangian, ${\cal L}_{\rm SM}$, is the most
general renormalizable Lagrangian that is consistent with the gauge
symmetry (\ref{smsym}), the particle content (\ref{ferrep}) and the
pattern of spontaneous symmetry breaking (\ref{smssb}). It can be
divided to three parts:
\beq\label{LagSM}
{\cal L}_{\rm SM}={\cal L}_{\rm kinetic}+{\cal L}_{\rm Higgs}
+{\cal L}_{\rm Yukawa}.
\eeq

As concerns the kinetic terms, to maintain gauge invariance, one has 
to replace the derivative with a covariant derivative:
\beq\label{SMDmu}
D^\mu=\partial^\mu+ig_s G^\mu_a L_a+ig W^\mu_b T_b+ig^\prime B^\mu Y.
\eeq
Here $G^\mu_a$ are the eight gluon fields, $W^\mu_b$ the three
weak interaction bosons and $B^\mu$ the single hypercharge boson.
The $L_a$'s are $SU(3)_{\rm C}$ generators (the $3\times3$
Gell-Mann matrices ${1\over2}\lambda_a$ for triplets, $0$ for singlets),
the $T_b$'s are $SU(2)_{\rm L}$ generators (the $2\times2$
Pauli matrices ${1\over2}\tau_b$ for doublets, $0$ for singlets),
and the $Y$'s are the $U(1)_{\rm Y}$ charges. For example, for the
quark doublets $Q_L$, we have
\beq\label{DmuQL}
{\cal L}_{\rm kinetic}(Q_L)= i{\overline{Q_{Li}}}\gamma_\mu
\left(\partial^\mu+{i\over2}g_s G^\mu_a\lambda_a
+{i\over2}g W^\mu_b\tau_b+{i\over6}g^\prime
B^\mu\right)\delta_{ij}Q_{Lj}, 
\eeq
while for the lepton doublets $L_L^I$, we have
\beq\label{DmuLL}
{\cal L}_{\rm kinetic}(L_L)= i{\overline{L_{Li}}}\gamma_\mu
\left(\partial^\mu+{i\over2}g W^\mu_b\tau_b-\frac i2 g^\prime
  B^\mu\right)\delta_{ij}L_{Lj}. 
\eeq
The unit matrix in flavor space, $\delta_{ij}$, signifies that 
these parts of the interaction Lagrangian are flavor-universal. In
addition, they conserve CP.

The Higgs potential, which describes the scalar self interactions, is given by:
\beq\label{HiPo}
{\cal L}_{\rm Higgs}=\mu^2\phi^\dagger\phi-\lambda(\phi^\dagger\phi)^2.
\eeq
For the Standard Model scalar sector, where there is a single doublet,
this part of the Lagrangian is also CP conserving.  

The quark Yukawa interactions are given by
\beq\label{Hqint}
-{\cal L}_{\rm Y}^{q}=Y^d_{ij}{\overline {Q_{Li}}}\phi D_{Rj}
+Y^u_{ij}{\overline {Q_{Li}}}\tilde\phi U_{Rj}+{\rm h.c.},
\eeq
(where $\tilde\phi=i\tau_2\phi^\dagger$) while the lepton Yukawa
interactions are given by 
\beq\label{Hlint}
-{\cal L}_{\rm Y}^{\ell}=Y^e_{ij}{\overline {L_{Li}}}\phi E_{Rj}
+{\rm h.c.}.
\eeq
This part of the Lagrangian is, in general, flavor-dependent (that is,
$Y^f\not\propto{\bf 1}$) and CP violating.

%%%%%%%%%%%%%
\subsection{Global symmetries and parameter counting}
\label{sec:spurions}
In the absence of the Yukawa matrices $Y^d$, $Y^u$ and $Y^e$, the SM
has a large $U(3)^5$ global symmetry:
\beq\label{gglobal}
G_{\rm global}(Y^{u,d,e}=0)=SU(3)_q^3\times SU(3)_\ell^2\times U(1)^5,
\eeq
where
\beqa\label{susuu}
SU(3)_q^3&=&SU(3)_Q\times SU(3)_U\times SU(3)_D,\no\\
SU(3)_\ell^2&=&SU(3)_L\times SU(3)_E,\no\\
U(1)^5&=&U(1)_B\times U(1)_L\times U(1)_Y\times U(1)_{\rm PQ}\times
U(1)_E.
\eeqa
Out of the five $U(1)$ charges, three can be identified with baryon
number ($B$), lepton number ($L$) and hypercharge ($Y$), which are 
respected by the Yukawa interactions. The two remaining $U(1)$ groups
can be identified with the PQ symmetry whereby the Higgs and $D_R,E_R$ 
fields have opposite charges, and with a global rotation of $E_R$
only. 

The Yukawa interactions (\ref{Hqint}) and (\ref{Hlint}) break the
global symmetry (of course, the gauged $U(1)_Y$ remains a good symmetry),
\beq\label{globre}
G_{\rm global}(Y^{u,d,e}\neq0)= U(1)_B\times U(1)_e\times
U(1)_\mu\times U(1)_\tau.
\eeq
One can think of the quark Yukawa couplings as spurions that break the
global $SU(3)_q^3$ symmetry (but are neutral under $U(1)_B$),
\beq\label{Gglobq}
Y^u\sim(3,\bar3,1)_{SU(3)_q^3},\ \ \ 
Y^d\sim(3,1,\bar3)_{SU(3)_q^3},
\eeq
and of the lepton Yukawa couplings as spurions that break the global
$SU(3)_\ell^2$ symmetry (but are neutral under $U(1)_e\times
U(1)_\mu\times U(1)_\tau$),  
\beq\label{Gglobl}
Y^e\sim(3,\bar3)_{SU(3)_\ell^2}.
\eeq
The spurion formalism is convenient for several purposes: parameter
counting (see below), identification of flavor suppression factors
(see Section \ref{sec:nppuzzle}), and the idea of minimal flavor
violation (see Section \ref{sec:lhc}).

How many independent parameters are there in ${\cal L}_{\rm Y}^q$? The
two Yukawa matrices, $Y^u$ and $Y^d$, are $3\times3$ and complex.
Consequently, there are 18 real and 18 imaginary parameters in these
matrices. Not all of them are, however, physical. The pattern of
$G_{\rm global}$ breaking means that there is freedom to remove 9 real
and 17 imaginary parameters (the number of parameters in three
$3\times3$ unitary matrices minus the phase related to $U(1)_B$). For
example, we can use the unitay transformations $Q_L\to V_QQ_L$,
$U_R\to V_U U_R$ and $D_R\to V_D D_R$, to lead to the following
interaction basis:
\beq\label{speint}
Y^d=\lambda_d,\ \ \ Y^u=V^\dagger\lambda_u,
\eeq
where $\lambda_{d,u}$ are diagonal,
\beq\label{deflamd}
\lambda_d={\rm diag}(y_d,y_s,y_b),\ \ \
\lambda_u={\rm diag}(y_u,y_c,y_t),
\eeq
while $V$ is a unitary matrix that depends on three real angles and
one complex phase. We conclude that there are 10 quark flavor
parameters: 9 real ones and a single phase. In the mass basis, we will
identify the nine real parameters as six quark masses and
three mixing angles, while the single phase is $\delta_{\rm KM}$.

How many independent parameters are there in ${\cal L}_{\rm Y}^\ell$?
The Yukawa matrix $Y^e$ is $3\times3$ and complex. Consequently, there
are 9 real and 9 imaginary parameters in this matrix. There is,
however, freedom to remove 6 real and 9 imaginary parameters (the
number of parameters in two $3\times3$ unitary matrices minus the
phases related to $U(1)^3$). For example, we can use the unitay
transformations $L_L\to V_LL_L$ and $E_R\to V_E E_R$, to lead to the
following interaction basis:
\beq\label{speintl}
Y^e=\lambda_e={\rm diag}(y_e,y_\mu,y_\tau). 
\eeq
We conclude that there are 3 real lepton flavor parameters. In the
mass basis, we will identify these parameters as the three
charged lepton masses. We must, however, modify the model when
we take into account the evidence for neutrino masses.

%%%%%%%%%%%%%%%%%%%
\subsection{The mass basis}
Upon the replacement $\re{\phi^0}\to\frac{v+H^0}{\sqrt2}$, the Yukawa
interactions (\ref{Hqint}) give rise to the mass matricess
\beq\label{YtoMq}
M_q={v\over\sqrt2}Y^q.
\eeq
The mass basis corresponds, by definition, to diagonal mass
matrices. We can  always find unitary matrices $V_{qL}$ and $V_{qR}$
such that 
\beq\label{diagMq}
V_{qL}M_q V_{qR}^\dagger=M_q^{\rm diag}\equiv\frac{v}{\sqrt2}\lambda_q.
\eeq
The four matrices $V_{dL}$, $V_{dR}$, $V_{uL}$ and $V_{uR}$ are then
the ones required to transform to the mass basis. For example, if we
start from the special basis (\ref{speint}), we have
$V_{dL}=V_{dR}=V_{uR}={\bf 1}$ and $V_{uL}=V$. The combination
$V_{uL}V_{dL}^\dagger$ is independent of the interaction basis from
which we start this procedure.

We denote the left-handed quark mass eigenstates as $U_L$ and $D_L$. 
The charged current interactions for quarks [that is the interactions of the 
charged $SU(2)_{\rm L}$ gauge bosons $W^\pm_\mu={1\over\sqrt2}
(W^1_\mu\mp iW_\mu^2)$], which in the interaction basis are described 
by (\ref{DmuQL}), have a complicated form in the mass basis:
\beq\label{Wmasq}
-{\cal L}_{W^\pm}^q={g\over\sqrt2}{\overline {U_{Li}}}\gamma^\mu
V_{ij}D_{Lj} W_\mu^++{\rm h.c.}.
\eeq
where $V$ is the $3\times3$ unitary matrix ($VV^\dagger=V^\dagger
V={\bf 1}$) that appeared in Eq. (\ref{speint}). For a general
interaction basis, 
\beq\label{VCKM}
V=V_{uL}V_{dL}^\dagger.
\eeq
$V$ is the Cabibbo-Kobayashi-Maskawa (CKM) {\it mixing matrix} for
quarks \cite{Cabibbo:1963yz,Kobayashi:1973fv}. As a result of the fact
that $V$ is not diagonal, the $W^\pm$ gauge bosons couple to quark
mass eigenstates of different generations. Within the Standard
Model, this is the only source of {\it flavor changing} quark
interactions.

{\bf Exercise 1:} {\it Prove that, in the absence of neutrino masses, there
is no mixing in the lepton sector.}

{\bf Exercise 2:} {\it Prove that there is no mixing in the $Z$
couplings. (In the physics jargon, there are no flavor changing
neutral currents at tree level.)}

The detailed structure of the CKM matrix, its parametrization, and the
constraints on its elements are described in Appendix \ref{app:ckm}.

%%%%%%%%%%%%%%%%%%%%%%%
\section{The new physics flavor puzzle}
\label{sec:nppuzzle}
It is clear that the Standard Model is not a complete theory of
Nature:
\begin{enumerate}
\item It does not include gravity, and therefore it cannot be valid at
  energy scales above $m_{\rm Planck}\sim10^{19}$ GeV:
\item It does not allow for neutrino masses, and therefore it cannot
  be valid at energy scales above $m_{\rm seesaw}\sim10^{15}$ GeV;
\item The fine-tuning problem of the Higgs mass and the puzzle of the
  dark matter suggest that the scale where the SM is replaced with a
  more fundamental theory is actually much lower, $\Lambda_{\rm
    NP}\lsim1$ TeV.
\end{enumerate}
Given that the SM is only an effective low energy theory,
non-renormalizable terms must be added to ${\cal L}_{\rm SM}$ of Eq.
(\ref{LagSM}). These are terms of dimension higher than four in the
fields which, therefore, have couplings that are inversely
proportional to the scale of new physics $\Lambda_{\rm NP}$. For
example, the lowest dimension non-renormalizable terms are dimension
five:
\beq\label{Hnint}
-{\cal L}_{\rm Yukawa}^{\rm dim-5}=
{Z_{ij}^\nu\over \Lambda_{\rm
    NP}}L_{Li}^I L_{Lj}^I\phi\phi+{\rm h.c.}. 
\eeq
These are the seesaw terms, leading to neutrino masses. We
will return to the topic of neutrino masses in section \ref{sec:nu}.

{\bf Exercise 3:} {\it How does the global symmetry breaking pattern
(\ref{globre}) change when (\ref{Hnint}) is taken into account?}

{\bf Exercise 4:} {\it What is the number of physical lepton flavor
parameters in this case? Identify these parameters in the mass basis.}

As concerns quark flavor physics, consider, for example, the following
dimension-six, four-fermion, flavor changing operators:
\beq
{\cal L}_{\Delta F=2}=
\frac{z_{sd}}{\Lambda_{\rm NP}^2}(\overline{d_L}\gamma_\mu s_L)^2
+\frac{z_{cu}}{\Lambda_{\rm NP}^2}(\overline{c_L}\gamma_\mu u_L)^2
+\frac{z_{bd}}{\Lambda_{\rm NP}^2}(\overline{d_L}\gamma_\mu b_L)^2
+\frac{z_{bs}}{\Lambda_{\rm NP}^2}(\overline{s_L}\gamma_\mu b_L)^2.
\eeq
Each of these terms contributes to the mass splitting between the
corresponding two neutral mesons. For example, the term ${\cal
  L}_{\Delta B=2}\propto(\overline{d_L}\gamma_\mu b_L)^2$ contributes
to $\Delta m_B$, the mass difference between the two neutral
$B$-mesons. We use $M_{12}^B=\frac{1}{2m_B}\langle B^0|{\cal
  L}_{\Delta F=2}|\overline{B}^0\rangle$ and
\beq
\langle B^0|(\overline{d_{La}}\gamma^\mu
b_{La})(\overline{d_{Lb}}\gamma_\mu b_{Lb})|\overline{B}^0\rangle =
-\frac13 m_B^2f_B^2 B_B.
\eeq
Analogous expressions hold for the other neutral mesons.\footnote{The
PDG \cite{Yao:2006px} quotes the following values, extracted from
leptonic charged meson decays: $f_K\approx0.16\ GeV$,
$f_D\approx0.23\ GeV$, $f_B\approx0.18\ GeV$. We further use
$f_{B_s}\approx0.20\ GeV$.}  This leads to $\Delta
m_B/m_B=2|M_{12}^B|/m_B\sim (z_{bd}/3)(f_B/\Lambda_{\rm NP})^2$.
Experiments give:
\beqa
\epsilon_K&\sim&2.3\times10^{-3},\no\\
\Delta m_K/m_K&\sim&7.0\times10^{-15},\no\\
\Delta m_D/m_D&\lsim&2\times10^{-14},\no\\
\Delta m_B/m_B&\sim&6.3\times10^{-14},\no\\
\Delta m_{B_s}/m_{B_s}&\sim&2.1\times10^{-12}.
\eeqa
These measurements give then the following constraints (the bound on
${\cal I}m(z_{sd})$ is stronger by a factor of
$(2\sqrt{2}\epsilon_K)^{-1}$ than the bound on $|z_{sd}|$):
\beq\label{lowlnp}
\Lambda_{\rm NP}\gsim\cases{
  \sqrt{{\cal I}m(z_{sd})}\ 2\times10^4\ TeV&$\epsilon_K$\cr
  \sqrt{z_{sd}}\ 1\times10^3\ TeV&$\Delta m_K$\cr
  \sqrt{z_{cu}}\ 9\times10^2\ TeV&$\Delta m_D$\cr
  \sqrt{z_{bd}}\ 4\times10^2\ TeV&$\Delta m_B$\cr
  \sqrt{z_{bs}}\ 7\times10^1\ TeV&$\Delta m_{B_s}$\cr}
\eeq
If the new physics has a generic flavor structure, that is
$z_{ij}={\cal O}(1)$, then its scale must be above $10^3-10^4$ TeV
(or, if the leading contributions involve electroweak loops, above
$10^2-10^3$ TeV). {\it If indeed $\Lambda_{\rm NP}\gg TeV$, it means
that we have misinterpreted the hints from the fine-tuning problem
and the dark matter puzzle.} There is, however, another way to look
at these constraints:
\beqa\label{zcons}
{\cal I}m(z_{sd})&\lsim&6\times10^{-9}\ (\Lambda_{\rm NP}/TeV)^2,\no\\
z_{sd}&\lsim&8\times10^{-7}\ (\Lambda_{\rm NP}/TeV)^2,\no\\
z_{cu}&\lsim&1\times10^{-6}\ (\Lambda_{\rm NP}/TeV)^2,\no\\
z_{bd}&\lsim&6\times10^{-6}\ (\Lambda_{\rm NP}/TeV)^2,\no\\
z_{bs}&\lsim&2\times10^{-4}\ (\Lambda_{\rm NP}/TeV)^2.
\eeqa
{\it It could be that the scale of new physics is of order TeV, but
  its flavor structure is far from generic.}

One can use that language of effective operators also for the SM,
integrating out all particles significantly heavier than the neutral
mesons (that is, the top, the Higgs and the weak gauge bosons). Thus,
the scale is $\Lambda_{\rm SM}\sim m_W$. Since the leading
contributions to neutral meson mixings come from box diagrams, the
$z_{ij}$ coefficients are suppressed by $\alpha_2^2$. To identify
the relevant flavor suppression factor, one can employ the spurion
formalism. For example, the flavor transition that is relevant to
$B^0-\overline{B}{}^0$ mixing involves $\overline{d_L}b_L$ which
transforms as $(8,1,1)_{SU(3)_q^3}$. The leading contribution must then
be proportional to $(Y^u Y^{u\dagger})_{13}\propto y_t^2
V_{tb}V_{td}^*$. Indeed, an explicit calculation (using VIA for the
matrix element and neglecting QCD corrections) gives\footnote{A detailed
derivation can be found in Appendix B of \cite{Branco:1999fs}.}
\beq \frac{2M_{12}^B}{m_B}\approx-\frac{\alpha_2^2}{12}
\frac{f_B^2}{m_W^2}S_0(x_t)(V_{tb}V_{td}^*)^2,
\eeq
where $x_i=m_i^2/m_W^2$ and
\beq
S_0(x)=\frac{x}{(1-x)^2}\left[1-\frac{11x}{4}+\frac{x^2}{4}-\frac{3x^2\ln
x}{2(1-x)}\right].  \eeq
Similar spurion analyses, or explicit calculations, allow us to
extract the weak and flavor suppression factors that apply in the
SM:
\beqa
{\cal I}m(z_{sd}^{\rm SM})&\sim&\alpha_2^2 y_t^2
|V_{td}V_{ts}|^2\sim1\times10^{-10},\no\\
z_{sd}^{\rm SM}&\sim&\alpha_2^2 y_c^2
|V_{cd}V_{cs}|^2\sim5\times10^{-9},\no\\  
z_{bd}^{\rm SM}&\sim&\alpha_2^2 y_t^2
|V_{td}V_{tb}|^2\sim7\times10^{-8},\no\\
z_{bs}^{\rm SM}&\sim&\alpha_2^2 y_t^2
|V_{ts}V_{tb}|^2\sim2\times10^{-6}.
\eeqa
(We did not include $z_{cu}^{\rm SM}$ in the list because it requires
a more detailed consideration. The naively leading short distance
contribution is $\propto \alpha_2^2(y_s^4/y_c^2)
|V_{cs}V_{us}|^2\sim5\times10^{-13}$. However, higher dimension terms
can replace a $y_s^2$ factor with $(\Lambda/m_D)^2$
\cite{Bigi:2000wn}. Moreover, long distance contributions are expected
to dominate. In particular, peculiar phase space effects
\cite{Falk:2001hx,Falk:2004wg} have been identified which are expected
to enhance $\Delta m_D$ to within an order of magnitude of the present
upper bound.)

It is clear than that contributions from new physics at $\Lambda_{\rm
  NP}\sim1\ TeV$ should be suppressed by factors that are comparable
or smaller than the SM ones. Why does that happen? This is the new
physics flavor puzzle.

The fact that the flavor structure of new physics at the TeV scale
must be non-generic means that flavor measurements are a good probe of
the new physics. Perhaps the best-studied example is that of
supersymmetry. Here, the spectrum of the superpartners and the
structure of their couplings to the SM fermions will allow us to probe
the mechanism of dynamical supersymmetry breaking.

%%%%%%%%%%%%%%%%%%%%%
\section{Lessons from $D^0-\overline{D}^0$ mixing}
\label{sec:dmix}
Interesting experimental results concerning $D^0-\overline{D}^0$
mixing have been recently achieved by the BELLE and BABAR experiments.
For the first time, there is evidence for width splitting (of order
one percent) between the two neutral $D$-mesons
\cite{Aubert:2007wf,Staric:2007dt}, while the bound on the mass splitting
has become stronger \cite{Abe:2007rd}. We use this recent experimental
information to draw important lessons on supersymmetry. This
demonstrates how flavor physics -- at the GeV scale -- provides a
significant probe of supersymmetry -- at the TeV scale.

%%%%%%%%%%%%
\subsection{Neutral meson mixing with supersymmetry}
We consider the contributions from the box diagrams involving the
squark doublets of the first two generations, $\tilde Q_{L1,2}$, to
the $D^0-\overline{D}^0$ and $K^0-\overline{K}^0$ mixing amplitudes.
The contributions that are relevant to the neutral $D$ system are
proportional to $K_{2i}^u K^{u*}_{1i}K_{2j}^u K^{u*}_{1j}$, where
$K^u$ is the mixing matrix of the gluino couplings to a left-handed up
quark and their supersymmetric squark partners. (In the language of
the mass insertion approximation, we calculate here the contribution
that is $\propto[(\delta^u_{LL})_{12}]^2$.) The contributions that are 
relevant to the neutral $K$ system are proportional to $K_{2i}^{d*}
K^{d}_{1i}K_{2j}^{d*} K^{d}_{1j}$, where $K^d$ is the mixing matrix of
the gluino couplings to a left-handed down quark and their
supersymmetric squark partners ($\propto[(\delta^d_{LL})_{12}]^2$ in
the mass insertion approximation). We work in the mass basis for both
quarks and squarks.
A detailed derivation \cite{Raz:2002zx} is given in Appendix
\ref{app:susyd}. It gives:
\beqa\label{motsusyb}
M_{12}^D&=&\frac{\alpha_s^2m_Df_D^2B_D\eta_{\rm QCD}}{108m_{\tilde u}^2}
[11\tilde f_6(x_u)+4x_uf_6(x_u)]\frac{(\Delta m^2_{\tilde
    u})^2}{m_{\tilde u}^4} (K_{21}^uK_{11}^{u*})^2,\\
\label{motsusyc}
M_{12}^K&=&\frac{\alpha_s^2m_Kf_K^2B_K\eta_{\rm QCD}}{108m_{\tilde d}^2}
[11\tilde f_6(x_d)+4x_df_6(x_d)]\frac{(\Delta\tilde m^2_{\tilde
    d})^2}{\tilde m_d^4} (K_{21}^{d*}K_{11}^{d})^2.
\eeqa
Here $m_{\tilde u,\tilde d}$ is the average mass of the corresponding
two squark generations, $\Delta m^2_{\tilde u,\tilde d}$ is the
mass-squared difference, and $x_{u,d}=m_{\tilde g}^2/m_{\tilde
  u,\tilde d}^2$.

One can immediately identify three generic ways in which
supersymmetric contributions to neutral meson mixing can be
suppressed: 
\begin{enumerate}
\item Heaviness: $m_{\tilde q}\gg1\ TeV$;
\item Degeneracy: $\Delta m^2_{\tilde q}\ll m_{\tilde q}^2$;
\item Alignment: $K^{d,u}_{21}\ll1$.
\end{enumerate}
When heaviness is the only suppression mechanism, as in split
supersymmetry \cite{ArkaniHamed:2004fb}, the squarks are very heavy
and supersymmetry no longer solves the fine tuning
problem.\footnote{When the first two squark generations are
mildly heavy and the third generation is light, as in effective
supersymmetry \cite{Cohen:1996vb}, the 
fine tuning problem is still solved, but additional suppression
mechanisms are needed.} If we want to maintain supersymmetry as a
solution to the fine tuning problem, either degeneracy or alignment or
a combination of both is needed. This means that the flavor structure
of supersymmetry is not generic, as argued in the previous section.

The $2\times2$ mass-squared matrices for the relevant squarks have the
following form:
\beqa\label{mllot}
\tilde M^2_{U_L}&=&\tilde m^2_{Q_L}+\left(\frac12-\frac23
  s^2_W\right)m_Z^2\cos2\beta+M_u M_u^\dagger,\no\\
\tilde M^2_{D_L}&=&\tilde m^2_{Q_L}-\left(\frac12-\frac13
  s^2_W\right)m_Z^2\cos2\beta+M_d M_d^\dagger.
\eeqa
We note the following features of the various terms:
\begin{itemize}
\item $\tilde m^2_{Q_L}$ is a $2\times2$ hermitian matrix of soft
supersymmetry breaking terms. It does not break $SU(2)_{\rm L}$ and
consequently it is common to $\tilde M^2_{U_L}$ and $\tilde M^2_{D_L}$. On the
other hand, it breaks in general the $SU(2)_Q$ flavor symmetry.
\item The terms proportional to $m_Z^2$ are the D-terms. They break
  supersymmetry (since they involve $D_{T_3}\neq0$ and for $D_Y\neq0$)
  and $SU(2)_{\rm L}$ but conserve $SU(2)_Q$.
\item The terms proportional to $M_q^2$ come from the $F_{U_R}$- and
  $F_{D_R}$-terms. They break the gauge $SU(2)_{\rm L}$ and the global
  $SU(2)_Q$ but, since $F_{U_R}=F_{D_R}=0$, conserve supersymmetry.
\end{itemize}
Given that we are interested in squark masses close to the TeV scale
(and the experimental lower bounds are of order 300 GeV), the scale of the 
eigenvalues of $\tilde m^2_{Q_L}$ is much higher than
$m_Z^2$ which, in turn, is much higher than $m_c^2$, the largest
eigenvalue in $M_q M_q^\dagger$. We can draw the following conclusions:
\begin{enumerate}
\item $m_{\tilde u}^2=m_{\tilde d}^2\equiv m_{\tilde q}^2$ up to
  effects of order $m_Z^2$, namely to an accuracy of ${\cal O}(10^{-2})$.
\item $\Delta m^2_{\tilde u}=\Delta m^2_{\tilde d}\equiv \Delta
  m^2_{\tilde q}$ up to effects of order $m_c^2$, namely to an
  accuracy of ${\cal O}(10^{-5})$.
\item Since $K_u\simeq V_{uL} \tilde V_L^\dagger$ and $K_d\simeq
  V_{dL} \tilde V_L^\dagger$ (the matrices $V_{qL}$ are defined in Eq.
  (\ref{diagMq}), while $\tilde V_L$ diagonalizes $\tilde m^2_{Q_L}$),
  the mixing matrices $K^u$ and $K^d$ are different from each other,
  but the following relation to the CKM matrix holds to an accuracy of
  ${\cal O}(10^{-5})$: 
\beq\label{kkckm}
K^u K^{d\dagger} = V. 
\eeq
\end{enumerate}

%%%%%%%%%%%%%%%%%%
\subsection{Non-degenerate squarks at the LHC?}
Eqs. (\ref{motsusyb}) and (\ref{motsusyc}) can be translated into our
generic language: 
\beqa
\Lambda_{\rm NP}&=&m_{\tilde q},\\
z_{cu}&=&z_{12}\sin^2\theta_u,\no\\
z_{sd}&=&z_{12}\sin^2\theta_d,\no\\
z_{12}&=&\frac{11\tilde f_6(x)+4x f_6(x)}{18}\alpha_s^2
\left(\frac{\Delta\tilde m_{\tilde q}^2}{m_{\tilde q}^2}\right)^2,
\eeqa
with Eq. (\ref{kkckm}) giving
\beq\label{kkckmb}
\sin\theta_u-\sin\theta_d\approx\sin\theta_c=0.23.
\eeq

We now ask the following question: Is it possible that the first two
generation squarks, $\tilde Q_{L1,2}$, are accessible to the LHC
($m_{\tilde q}\lsim1\ TeV$), and are not degenerate ($\Delta
m^2_{\tilde q}/m_{\tilde q}^2={\cal O}(1)$)?

To answer this question, we use Eqs. (\ref{zcons}). For $\Lambda_{\rm
  NP}\lsim1\ TeV$, we have $z_{cu}\lsim1\times10^{-6}$ and, for a
phase that is $\not\ll0.1$, $z_{sd}\lsim6\times10^{-8}$. On the other
hand, for non-degenerate squarks, and, for example, $11\tilde
f_6(1)+4f_6(1)=1/6$, we have $z_{12}=8\times10^{-5}$. Then we need,
simultaneously, $\sin\theta_u\lsim0.11$ and
$\sin\theta_d\lsim0.03$, but this is inconsistent with Eq. (\ref{kkckmb}).

There are three ways out of this situation:
\begin{enumerate}
\item The first two generation squarks are quasi-degenerate. The minimal
level of degeneracy is $(\tilde m_2-\tilde m_1)/(\tilde m_2+\tilde
m_1)\lsim0.12$. It could be the result of RGE \cite{Nir:2002ah}.
\item The first two generation squarks are heavy. Putting
$\sin\theta_u=0.23$ and $\sin\theta_d\approx0$,
as in models of alignment \cite{Nir:1993mx,Leurer:1993gy}, Eq.
(\ref{lowlnp}) leads to
\beq\label{mqali}
m_{\tilde q}\gsim2\ TeV.
\eeq
\item The ratio $x=\tilde m_g^2/\tilde m_q^2$ is in a fine-tuned region of
parameter space where there are accidental cancellations in $11\tilde
f_6(x)+4xf_6(x)$. For example, for $x=2.33$, this combination is
$\sim0.003$ and the bound (\ref{mqali}) is relaxed by a factor of 7.
\end{enumerate}
Barring such accidental cancellations, the {\it model independent}
conclusion is that, if the first two generations of squark doublets
are within the reach of the LHC, they must be quasi-degenerate
\cite{Ciuchini:2007cw,Nir:2007ac}.

{\bf Exercise 5:} {\it Does $K_{31}^d\sim|V_{ub}|$ suffice to satisfy the
$\Delta m_B$ constraint with neither degeneracy nor heaviness? (Use
the two generation approximation and ignore the second generation.)}

Is there a natural way to make the squarks degenerate? Examining
Eqs. (\ref{mllot}) we learn that degeneracy requires $\tilde
m^2_{Q_L}\simeq\tilde m^2_{\tilde q}{\bf 1}$. We have mentioned
already that flavor universality is a generic feature of gauge
interactions. Thus, the requirement of degeneracy is perhaps a hint
that supersymmetry breaking is {\it gauge mediated} to the MSSM
fields.

%%%%%%%%%%%
%%%%%%%%%
\section{Flavor at the LHC}
\label{sec:lhc}
The LHC will study the physics of electroweak symmetry breaking. There
are high hopes that it will discover not only the Higgs, but also shed
light on the fine-tuning problem that is related to the Higgs mass.
Here, we focus on the issue of how, through the study of new physics,
the LHC can shed light on the new physics flavor puzzle.

%%%%%%%%%%
\subsection{Minimal flavor violation (MFV)}
If supersymmetry breaking is gauge mediated, the squark mass matrices
of Eq. (\ref{mllot}) have the following form:
\beqa\label{mllgm}
\tilde M^2_{U_L}&=&\left[m^2_{\tilde Q_L}+\left(\frac12-\frac23
  s^2_W\right)m_Z^2\cos2\beta\right]{\bf 1}+M_u M_u^\dagger,\no\\
\tilde M^2_{D_L}&=&\left[m^2_{\tilde Q_L}-\left(\frac12-\frac13
  s^2_W\right)m_Z^2\cos2\beta\right]{\bf 1}+M_d M_d^\dagger.
\eeqa
Here, and in all other squark mass matrices, the only source of the
$SU(3)^3_q$ breaking are the SM Yukawa matrices. 

Models of gauge mediated supersymmetry breaking (GMSB) provide a
concrete example of a large class of models that obey a simple
principle called {\it minimal flavor violation} (MFV)
\cite{D'Ambrosio:2002ex}. This principle guarantees that low energy
flavor changing processes deviate only very little from the SM
predictions.  The basic idea can be described as follows. The gauge
interactions of the SM are universal in flavor space. The only
breaking of this flavor universality comes from the three Yukawa
matrices, $Y_U$, $Y_D$ and $Y_E$. If this remains true in the presence
of the new physics, namely $Y_U$, $Y_D$ and $Y_E$ are the only flavor
non-universal parameters, then the model belongs to the MFV class.

Let us now formulate this principle in a more formal way, using the
language of spurions that we presented in section \ref{sec:spurions}.  
The Standard Model with vanishing Yukawa couplings has a large global
symmetry (\ref{gglobal},\ref{susuu}). In this section we concentrate
only on the quarks. The non-Abelian part of the flavor symmetry for
the quarks is $SU(3)_q^3$ of Eq. (\ref{susuu}) with  
the three generations of quark fields transforming as follows:
\beq
Q_L(3,1,1),\ \ U_R(1,3,1),\ \ D_R(1,1,3).
\eeq
The Yukawa interactions,
\beq\label{eq:lagy}
{\cal L}_Y=\overline{Q_L}Y_D D_R H + \overline{Q_L}Y_U U_R H_c ,
\eeq
($H_c=i\tau_2 H^*$) break this symmetry. The Yukawa couplings can thus
be thought of as spurions with the following transformation
properties under $SU(3)_q^3$ [see Eq. (\ref{Gglobq})]:
\beq
Y_U\sim(3,\bar3,1),\qquad Y_D\sim(3,1,\bar3).
\eeq
When we say ``spurions'', we mean that we pretend that the Yukawa
matrices are fields which transform under the flavor symmetry, and
then require that all the Lagrangian terms, constructed
from the SM fields, $Y_{D}$ and $Y_U$, must be (formally)
invariant under the flavor group $SU(3)_q^3$. Of course, in reality,
${\cal L}_Y$ breaks $SU(3)_q^3$ precisely because $Y_{D,U}$ are {\it
  not} fields and do not transform under the symmetry.

The idea of minimal flavor violation is relevant to extensions of the
SM, and can be applied in two ways:
\begin{enumerate}
 \item If we consider the SM as a low energy effective theory, then
   all higher-dimension operators, constructed from SM-fields and
   $Y$-spurions, are formally invariant under $G_{\rm global}$.
\item If we consider a full high-energy theory that extends the SM,
  then all operators, constructed from SM and the new fields, and from
  $Y$-spurions, are formally invariant under $G_{\rm global}$.
\end{enumerate}

{\bf Exercise 8:} {\it Use the spurion formalism to argue that, in MFV
models, the $K_L\to\pi^0\nu\bar\nu$ decay amplitude is proportional to
$y_t^2 V_{td}V_{ts}^*$.}

Examples of MFV models include models of supersymmetry with
gauge-mediation or with anomaly-mediation of its breaking.
If the LHC discovers new particles that couple to the SM fermions,
then it will be able to test solutions to the new physics flavor
puzzle such as MFV \cite{Grossman:2007bd}. Much of its power to test
such frameworks is based on identifying top and bottom quarks.

To understand this statement, we notice that the spurions $Y_U$ and
$Y_D$ can always be written in terms of the two diagonal Yukawa
matrices $\lambda_u$ and $\lambda_d$ and the CKM matrix $V$, see
Eqs. (\ref{speint},\ref{deflamd}). Thus, the only source of quark
flavor changing transitions in MFV models is the CKM matrix. Next,
note that to an accuracy that is better than ${\cal O}(0.05)$, we can 
write the CKM matrix as follows:
\beq\label{ckmapp}
V=\pmatrix{1&0.23&0\cr -0.23&1&0\cr 0&0&1\cr}.
\eeq

{\bf Exercise 9:} {\it The approximation (\ref{ckmapp}) should be 
intuitively obvious to top-physicists, but definitely
counter-intuitive to bottom-physicists. (Some of them have dedicated a
large part of their careers to experimental or theoretical efforts to
determine $V_{cb}$ and $V_{ub}$.) What does the approximation
imply for the bottom quark? When we take into account that it is
only good to ${\cal O}(0.05)$, what would the implications be?}

We learn that the third generation of quarks is decoupled, to a good
approximation, from the first two. This, in turn, means that any new
particle that couples to the SM quarks (think, for example, of heavy
quarks in vector-like representations of $G_{\rm SM}$), decay into
either third generation quark, or to non-third generation quark, but
not to both. For example, in Ref. \cite{Grossman:2007bd}, MFV models
with additional charge $-1/3$, $SU(2)_{\rm L}$-singlet quarks --
$B^\prime$ -- were considered. A concrete test of MFV was proposed,
based on the fact that the largest mixing effect involving the third
generation is of order $|V_{cb}|^2\sim0.002$: Is the following
prediction, concerning events of $B^\prime$ pair production,
fulfilled:
\beq
\frac{\Gamma(B^\prime\overline{B^\prime}\to Xq_{1,2}q_3)}
{\Gamma(B^\prime\overline{B^\prime}\to Xq_{1,2}q_{1,2})+
  \Gamma(B^\prime\overline{B^\prime}\to Xq_3q_3)}\lsim10^{-3}.
\eeq
If not, then MFV is excluded.

One can think of analogous tests in the supersymmetric framework
\cite{nish}. Here, there is also a generic prediction that, in each
sector ($Q_L,U_R,D_R$), squarks of the first two generations are
quasi-degenerate, and do not decay into third generation quarks.
Squarks of the third generation can be separated in mass (though, for
small $\tan\beta$, the degeneracy in the $\tilde D_R$ sector is
threefold), and decay only to third generation quarks.

We conclude that measurements at the LHC related to new particles that
couple to the SM fermions are likely to teach us much more about flavor
physics.

%%%%%%%%%%%%%%%%%%%%
%%%%%%%%%%%%%%%%%%%%%
\section{Lessons from $S_{\psi K_S}$}
\label{sec:bmix}
Measurements of rates, mixing, and CP asymmetries in $B$ decays in the
two B factories, BaBar abd Belle, and in the two Tevatron detectors,
CDF and D0, signified a new era in our understanding of CP
violation. The progress is both qualitative and quantitative. Various
basic questions concerning CP and flavor violation have received, for
the first time, answers based on experimental information. These
questions include, for example,
\begin{itemize}
\item Is the Kobayashi-Maskawa mechanism at work (namely, is
  $\delta_{\rm KM}\neq0$)?
\item Does the KM phase dominate the observed CP violation?
\end{itemize}
As a first step, one may assume the SM and test the overall
consistency of the various measurements. However, the richness of data
from the B factories allow us to go a step further and answer these
questions model independently, namely allowing new physics to
contribute to the relevant processes. We here explain the way in which
this analysis proceeds.

%%%%%%%%%%%%%%
\subsection{$S_{\psi K_S}$}
The CP asymmetry in $B\to\psi K_S$ decays plays a major role in
testing the KM mechanism. Before we explain the test itself, we should
understand why is the theoretical interpretation of the asymmetry
exceptionally clean, and what are the theoretical parameters on which
it depends, within and beyond the Standard Model.

The CP asymmetry in neutral meson decays into final CP eigenstates
$f_{\CP}$ is defined as follows:
\beq\label{asyfcpt}
{\cal A}_{f_{\CP}}(t)\equiv\frac{d\Gamma/dt[\Bzb_{\rm phys}(t)\to f_{\CP}]-
d\Gamma/dt[\Bz_{\rm phys}(t)\to f_{\CP}]}
{d\Gamma/dt[\Bzb_{\rm phys}(t)\to f_{\CP}]+d\Gamma/dt[\Bz_{\rm phys}(t)\to
  f_{\CP}]}\; .
\eeq
A detailed evaluation of this asymmetry is given in Appendix
\ref{sec:formalism}. It leads to the following form:
\beqa\label{asyfcpbt}
{\cal A}_{f_{\CP}}(t)&=&S_{f_{\CP}}\sin(\Delta
mt)-C_{f_{\CP}}\cos(\Delta mt),\no\\ 
S_{f_{\CP}}&\equiv&\frac{2\,\im{\lambda_{f_{\CP}}}}{1+|\lambda_{f_{\CP}}|^2},\
\ \  C_{f_{\CP}}\equiv\frac{1-|\lambda_{f_{\CP}}|^2}{1+|\lambda_{f_{\CP}}|^2}
\; ,
\eeqa
where
\beq\label{lamhad}
\lambda_{f_{\CP}}=e^{-i\phi_B}(\overline{A}_{f_{\CP}}/A_{f_{\CP}}) \; .
\eeq
Here $\phi_B$ refers to the phase of $M_{12}$ [see
Eq.~(\ref{defmgam})].  Within the Standard Model, the corresponding
phase factor is given by
\beq\label{phimsm}
e^{-i\phi_B}=(V_{tb}^* V_{td}^{})/(V_{tb}^{}V_{td}^*) \;.
\eeq
The decay amplitudes $A_f$ and $\overline{A}_f$ are defined in
Eq. (\ref{decamp}). 

\begin{figure}[htb]
\caption{Feynman diagrams for (a) tree and (b) penguin amplitudes
  contributing to $B^0\to f$ or $B_{s}\to f$ via a $\bar b\to\bar q
  q\bar q^\prime$ quark-level process.}
\label{fig:diags}
\begin{center}
\includegraphics[width=2.85in]{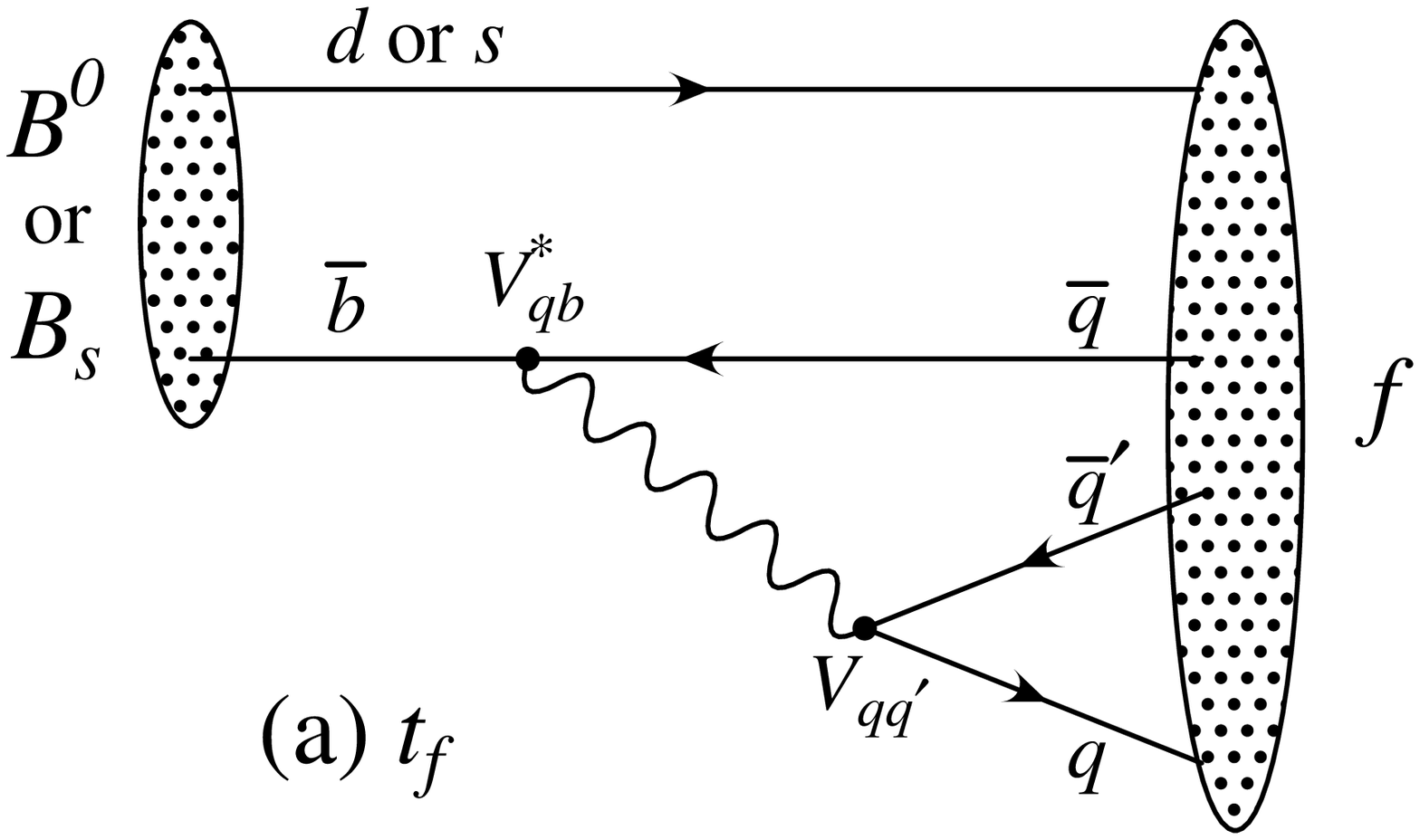}
\hspace{2em}
\includegraphics[width=2.85in]{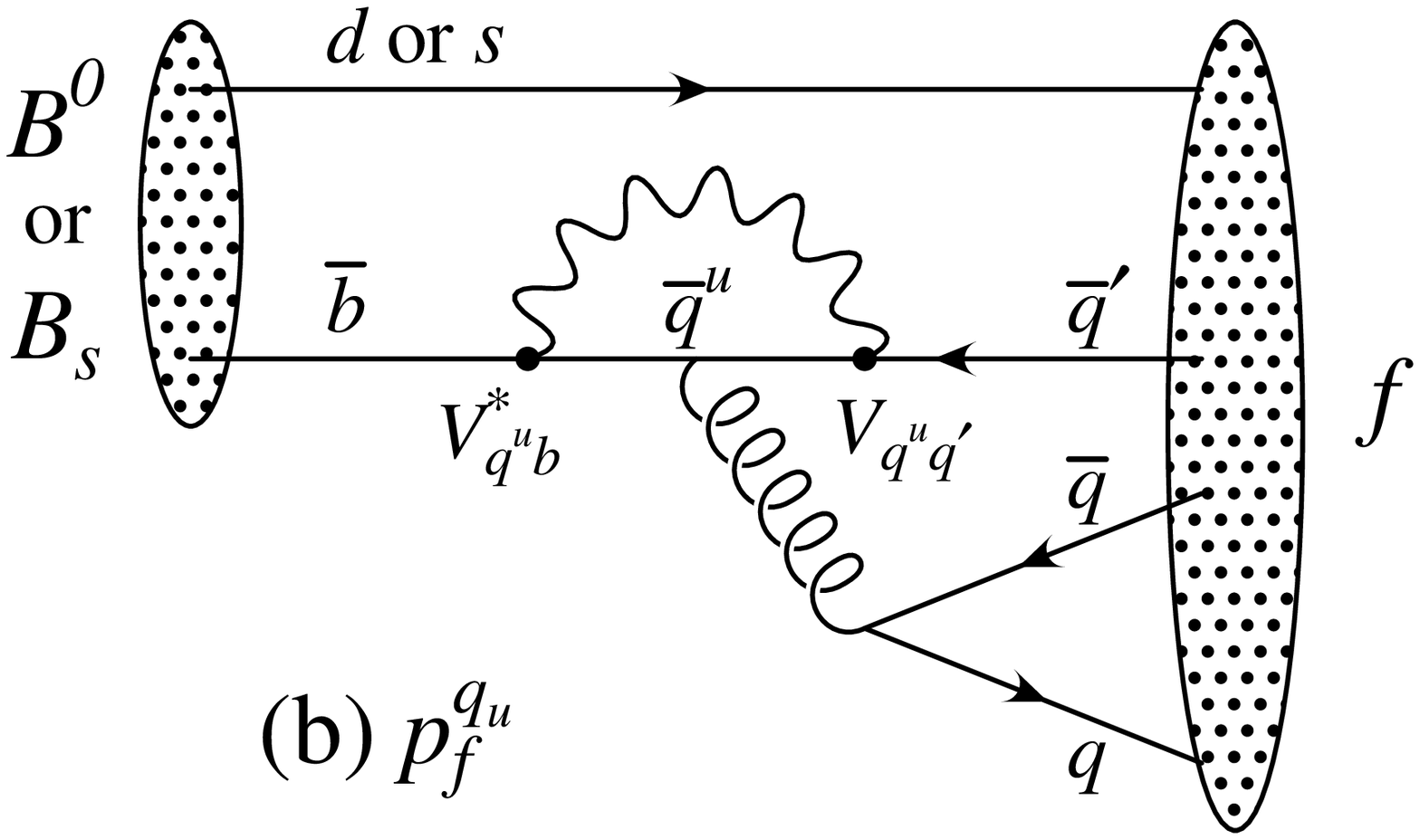}
\end{center}
\end{figure}

The $B^0\to J/\psi K^0$ decay~\cite{Carter:1980hr,Bigi:1981qs} proceeds
via the quark transition $\bar b\to\bar c c\bar s$. There are
contributions from both tree ($t$) and penguin ($p^{q_u}$, where
$q_u=u,c,t$ is the quark in the loop) diagrams (see
Fig.~\ref{fig:diags}) which carry different weak phases:
\beq\label{ckmdec}
A_f = \left(V^\ast_{cb} V^{}_{cs}\right) t_f +
\sum_{q_u= u,c,t}\left(V^\ast_{q_u b} V^{}_{q_u s}\right) p^{q_u}_f \; .
\eeq
(The distinction between tree and penguin contributions is a heuristic
one, the separation by the operator that enters is more precise. For a
detailed discussion of the more complete operator product approach,
which also includes higher order QCD corrections, see, for example,
ref. \cite{Buchalla:1995vs}.)  Using CKM unitarity, these decay
amplitudes can always be written in terms of just two CKM
combinations:
\beq\label{btoccs}
A_{\psi K}=\left(V^\ast_{cb} V^{}_{cs}\right)T_{\psi
  K}+\left(V^\ast_{ub} V^{}_{us}\right)P^u_{\psi K}, 
\eeq
where $T_{\psi K}=t_{\psi K}+p^c_{\psi K}-p^t_{\psi K}$ and
$P^u_{\psi K}=p^u_{\psi K}-p^t_{\psi K}$. A subtlety arises in this
decay that is related to the fact that ${B}^0\to J/\psi K^0$ and
$\overline{B}^0\to J/\psi\overline{K}{}^0$. A common final state,
e.g. $J/\psi K_S$, is reached only via $K^0-\overline{K}{}^0$ mixing. 
Consequently, the phase factor corresponding to neutral $K$ mixing,
$e^{-i\phi_K}=(V^*_{cd}V^{}_{cs})/(V^{}_{cd}V^*_{cs})$, plays a
role: 
\beq\label{psikmix}
\frac{\overline{A}_{\psi K_S}}{A_{\psi K_S}}
=-\frac{\left(V^{}_{cb} V^\ast_{cs}\right)T_{\psi
    K}+\left(V^{}_{ub} V^\ast_{us}\right)P^u_{\psi K}}
{\left(V^\ast_{cb} V^{}_{cs}\right)T_{\psi
    K}+\left(V^\ast_{ub} V^{}_{us}\right)P^u_{\psi K}}\times
\frac{V_{cd}^\ast V_{cs}^{}}{V_{cd}^{}V_{cs}^\ast}.
\eeq

The crucial point is that, for $B\to J/\psi K_S$ and other $\bar
b\to\bar cc\bar s$ processes, we can neglect the $P^u$ contribution to
$A_{\psi K}$, in the SM, to an approximation that is better than one
percent:
\beq\label{smapprox}
|P^u_{\psi K}/T_{\psi K}|\times|V_{ub}/V_{cb}|\times|
V_{us}/V_{cs}|\sim({\rm loop\ factor})\times0.1\times0.23\lsim0.005.
\eeq
Thus, to an accuracy of better than one percent,
\beq
\lambda_{\psi K_S}=\left(\frac{V_{tb}^*
  V_{td}^{}}{V_{tb}^{}V_{td}^*}\right)\left(\frac{V_{cb}
  V_{cd}^{*}}{V_{cb}^{*}V_{cd}}\right)=-e^{-2i\beta},
\eeq
where $\beta$ is defined in Eq. (\ref{abcangles}), and consequently
\beq\label{btopsik}
S_{\psi K_S}=\sin2\beta,\ \ \ C_{\psi K_S}=0 \; .
\eeq
(Below the percent level, several effects modify this equation
\cite{Grossman:2002bu,Boos:2004xp}.)

{\bf Exercise 6:} {\it Show that, if the $B\to\pi\pi$ decays were dominated 
by tree diagrams, then $S_{\pi\pi}=\sin2\alpha$.}

{\bf Exercise 7:} {\it Estimate the accuracy of the predictions
$S_{\phi K_S}=\sin2\beta$ and $C_{\phi K_S}=0$.}

When we consider extensions of the SM, we still do not expect any
significant new contribution to the tree level decay, $b\to c\bar cs$,
beyond the SM $W$-mediated diagram. Thus, the expression $\bar A_{\psi
  K_S}/A_{\psi K_S}=(V_{cb}V_{cd}^*)/(V_{cb}^*V_{cd})$ remains valid,
though the approximation of neglecting sub-dominant phases can be
somewhat less accurate than Eq. (\ref{smapprox}). On the other hand,
$M_{12}$, the $B^0-\overline{B}^0$ mixing amplitude, can in principle
get large and even dominant contributions from new physics. We can
parametrize the modification to the SM in terms of two parameters,
$r_d^2$ signifying the change in magnitude, and $2\theta_d$
signifying the change in phase:
\beq\label{derthed}
M_{12}=r_d^2\ e^{2i\theta_d}\ M_{12}^{\rm SM}(\rho,\eta).
\eeq
This leads to the following generalization of Eq. (\ref{btopsik}):
\beq\label{btopsiknp}
S_{\psi K_S}=\sin(2\beta+2\theta_d),\ \ \ C_{\psi K_S}=0 \; .
\eeq

The experimental measurements give the following ranges \cite{hfag}:
\beq\label{scpkexp}
S_{\psi K_S}=0.68\pm0.03,\ \ \ C_{\psi K_S}=0.01\pm0.02 \; .
\eeq

%%%%%%%%%%%%%%%
\subsection{Self-consistency of the CKM assumption}
The three generation standard model has room for CP violation, through
the KM phase in the quark mixing matrix. Yet, one would like to make
sure that indeed CP is violated by the SM interactions, namely that
$\sin\delta_{\rm KM}\neq0$. If we establish that this is the case, we
would further like to know whether the SM contributions to CP
violating observables are dominant. More quantitatively, we would like
to put an upper bound on the ratio between the new physics and the SM
contriubtions.

\begin{figure}[tb]
  \centering
  {\includegraphics[width=0.65\textwidth]{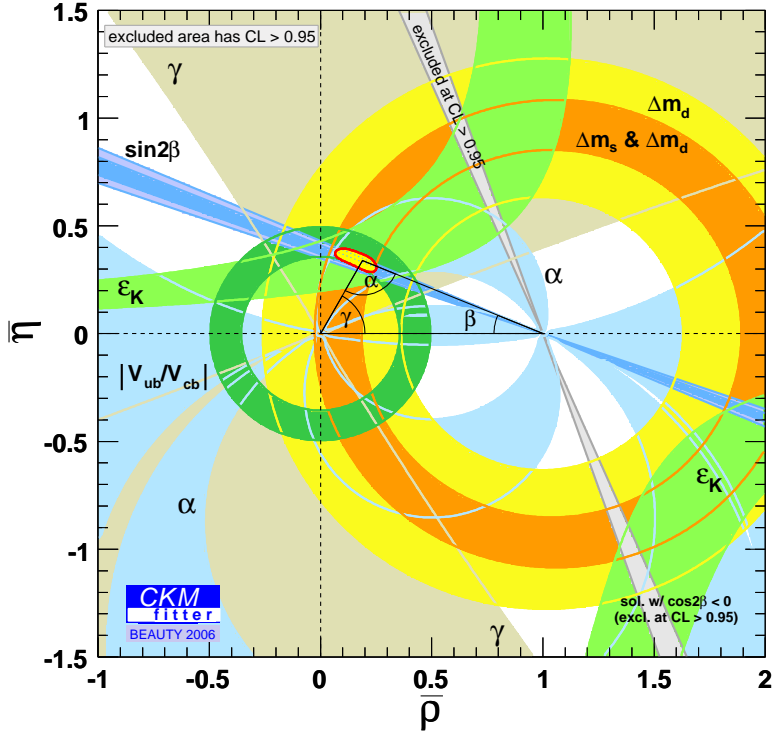}}
  \caption{Allowed region in the $\rho,\eta$ plane. Superimposed are
  the individual constraints from charmless semileptonic $B$ decays
  ($|V_{ub}/V_{cb}|$), mass differences in the $B^0$ ($\Delta m_d$)
  and $B_s$ ($\Delta m_s$) neutral meson systems, and CP violation in
  $K\to\pi\pi$ ($\varepsilon_K$), $B\to\psi K$ ($\sin2\beta$), 
  $B\to\pi\pi,\rho\pi,\rho\rho$ ($\alpha$), and $B\to DK$
  ($\gamma$). Taken from \cite{ckmfitter}.}
  \label{fg:UT}
\end{figure}

As a first step, one can assume that flavor changing processes are
fully described by the SM, and check the consistency of the various
measurements with this assumption. There are four relevant mixing
parameters, which can be taken to be the Wolfenstein parameters
$\lambda$, $A$, $\rho$ and $\eta$ defined in Eq. (\ref{wolpar}). The
values of $\lambda$ and $A$ are known rather accurately
\cite{Yao:2006px}: 
\beq\label{lamaexp}
\lambda=0.2272\pm0.0010,\ \ \ A=0.818^{+0.007}_{-0.017}.
\eeq
Then, one can express all the relevant observables as a function of
the two remaining parameters, $\rho$ and $\eta$, and check whether
there is a range in the $\rho-\eta$ plane that is consistent with all
measurements. The list of observables includes the following:
\begin{itemize}
\item The rates of inclusive and exclusive charmless semileptonic $B$
  decays depend on $|V_{ub}|^2\propto\rho^2+\eta^2$;
\item The CP asymmetry in $B\to\psi K_S$, $S_{\psi K_S}=\sin2\beta$
  with $e^{i\beta}=1-\rho+i\eta$;
\item The rates of various $B\to DK$ decays depend on the phase
  $\gamma$, where $e^{i\gamma}=\rho+i\eta$;
\item The rates of various $B\to\pi\pi,\rho\pi,\rho\rho$ decays depend
  on the phase $\alpha=\pi-\beta-\gamma$;
\item The ratio between the mass splittings in the neutral $B$ and
  $B_s$ systems is sensitive to $|V_{td}/V_{ts}|^2=(1-\rho)^2+\eta^2$;
\item The CP violation in $K\to\pi\pi$ decays, $\epsilon_K$, depends
  in a complicated way on $\rho$ and $\eta$.
\end{itemize}
The resulting constraints are shown in Fig. \ref{fg:UT}.
            
The consistency of the various constraints is impressive. In
particular, the following ranges for $\rho$ and $\eta$ can account for
all the measurements \cite{Yao:2006px}:
\beq
\rho=0.221^{+0.064}_{-0.028},\ \ \ \eta=0.340^{+0.017}_{-0.045}.
\eeq

One can make then the following statement \cite{Nir:2002gu}:\\
{\bf Very likely, CP violation in flavor changing processes is
  dominated by the Kobayashi-Maskawa phase.}

In the next two subsections, we explain how we can remove the phrase
``very likely'' from this statement, and how we can quantify the
KM-dominance.

%%%%%%%%%%%%%%%
\subsection{Is the KM mechanism at work?}
In proving that the KM mechanism is at work, we assume that
charged-current tree-level processes are dominated by the $W$-mediated
SM diagrams. This is a very plausible assumption. I am not aware of
any viable well-motivated model where this assumption is not valid.
Thus we can use all tree level processes and fit them to $\rho$ and
$\eta$, as we did before. The list of such processes includes the
following:
\begin{enumerate}
\item Charmless semileptonic $B$-decays, $b\to u\ell\nu$, measure
  $R_u$ [see Eq. (\ref{RbRt})].
\item $B\to DK$ decays, which go through the quark transitions $b\to
  c\bar u s$ and $b\to u\bar cs$, measure the angle $\gamma$ [see Eq.
  (\ref{abcangles})].
\item $B\to\rho\rho$ decays (and, similarly, $B\to\pi\pi$ and
  $B\to\rho\pi$ decays) go through the quark transition $b\to u\bar
  ud$. With an isospin analysis, one can determine the relative phase
  between the tree decay amplitude and the mixing amplitude. By
  incorporating the measurement of $S_{\psi K_S}$, one can subtract
  the phase from the mixing amplitude, finally providing a measurement
  of the angle $\gamma$ [see Eq. (\ref{abcangles})].
  \end{enumerate}

In addition, we can use loop processes, but then we must allow for new
physics contributions, in addition to the $(\rho,\eta)$-dependent SM
contributions. Of course, if each such measurement adds a separate 
mode-dependent parameter, then we do not gain anything by using this
information. However, there is a number of observables where the only
relevant loop process is $B^0-\overline{B}{}^0$ mixing. The list
includes $S_{\psi K_S}$, $\Delta m_B$ and the CP asymmetry in
semileptonic $B$ decays:
\beqa\label{apksNP}
S_{\psi K_S}&=&\sin(2\beta+2\theta_d),\no\\
\Delta m_{B}&=&r_d^2(\Delta m_B)^{\rm SM},\no\\
{\cal A}_{\rm SL}&=&-{\cal
    R}e\left(\frac{\Gamma_{12}}{M_{12}}\right)^{\rm
    SM}\frac{\sin2\theta_d}{r_d^2}
  +{\cal I}m\left(\frac{\Gamma_{12}}{M_{12}}\right)^{\rm
    SM}\frac{\cos2\theta_d}{r_d^2}.
\eeqa
As explained above, such process involve two new parameters [see Eq.
(\ref{derthed})]. Since there are three relevant observables, we can
further tighten the constraints in the $(\rho,\eta)$-plane. Similarly,
one can use measurements related to $B_s-\overline{B}_s$ mixing. One
gains three new observables at the cost of two new parameters (see,
for example, \cite{Grossman:2006ce}).

\begin{figure}[tb]
  \centering
  {\includegraphics[width=0.65\textwidth]{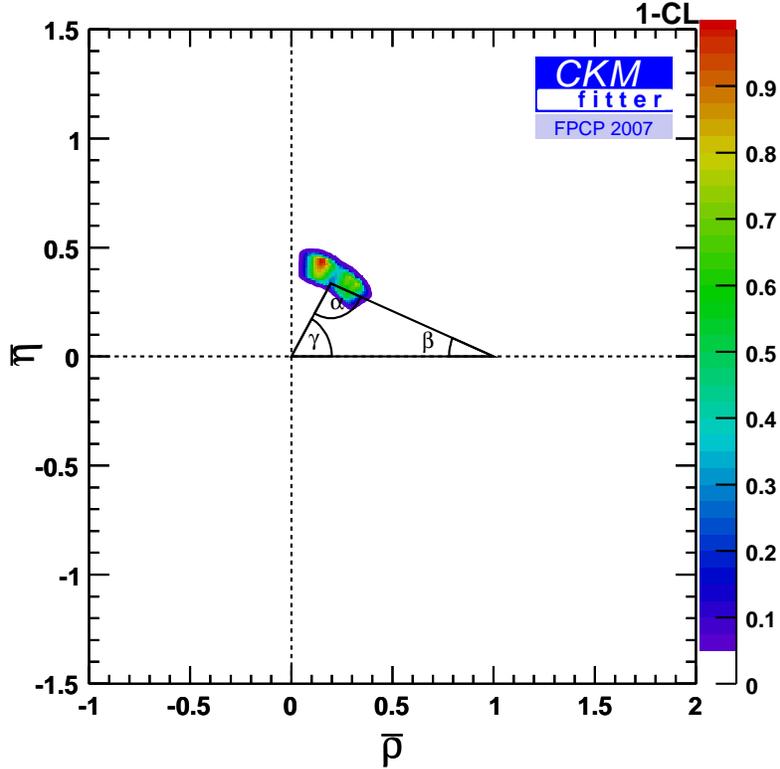}}
  \caption{The allowed region in the $\rho-\eta$ plane, assuming that
  tree diagrams are dominated by the Standard Model \cite{ckmfitter}.}
  \label{fig:re_tree}
\end{figure}

The results of such fit, projected on the $\rho-\eta$ plane, can be
seen in Fig. \ref{fig:re_tree}. It gives \cite{ckmfitter}
\beq
\eta=0.44^{+0.05}_{-0.23}\ \ (3\sigma).
\eeq
[A similar analysis in Ref. \cite{Bona:2007vi} obtains the $3\sigma$
range $(0.31-0.46)$.] It is clear that $\eta\neq0$ is well
established:\\
{\bf The Kobayashi-Maskawa mechanism of CP violation is at work.}

Another way to establish that CP is violated by the CKM matrix is to
find, within the same procedure, the allowed range for $\sin2\beta$
\cite{Bona:2007vi}: 
\beq\label{stbth}
\sin2\beta^{\rm tree}=0.76\pm0.04.
\eeq
(Ref. \cite{ckmfitter} finds $0.82^{+0.02}_{-0.13}$.) Thus,
$\beta\neq0$ is well established.

The consistency of the experimental results (\ref{scpkexp}) with the
SM predictions (\ref{btopsik},\ref{stbth}) means that the KM mechanism
of CP violation dominates the observed CP violation. In the next
subsection, we make this statement more quantitative.

%%%%%%%%%%%%%%%%%
\subsection{How much can new physics contribute to
  $B^0-\overline{B}{}^0$ mixing?}
\begin{figure}[htb]
\caption{Constraints in the (a) $r_d^2-2\theta_d$ plane, and (b)
  $h_d-\sigma_d$ plane, assuming that NP contributions to tree
  level processes are negligible \cite{ckmfitter}.}
\label{fig:rdtd}
\begin{center}
\includegraphics[width=2.85in]{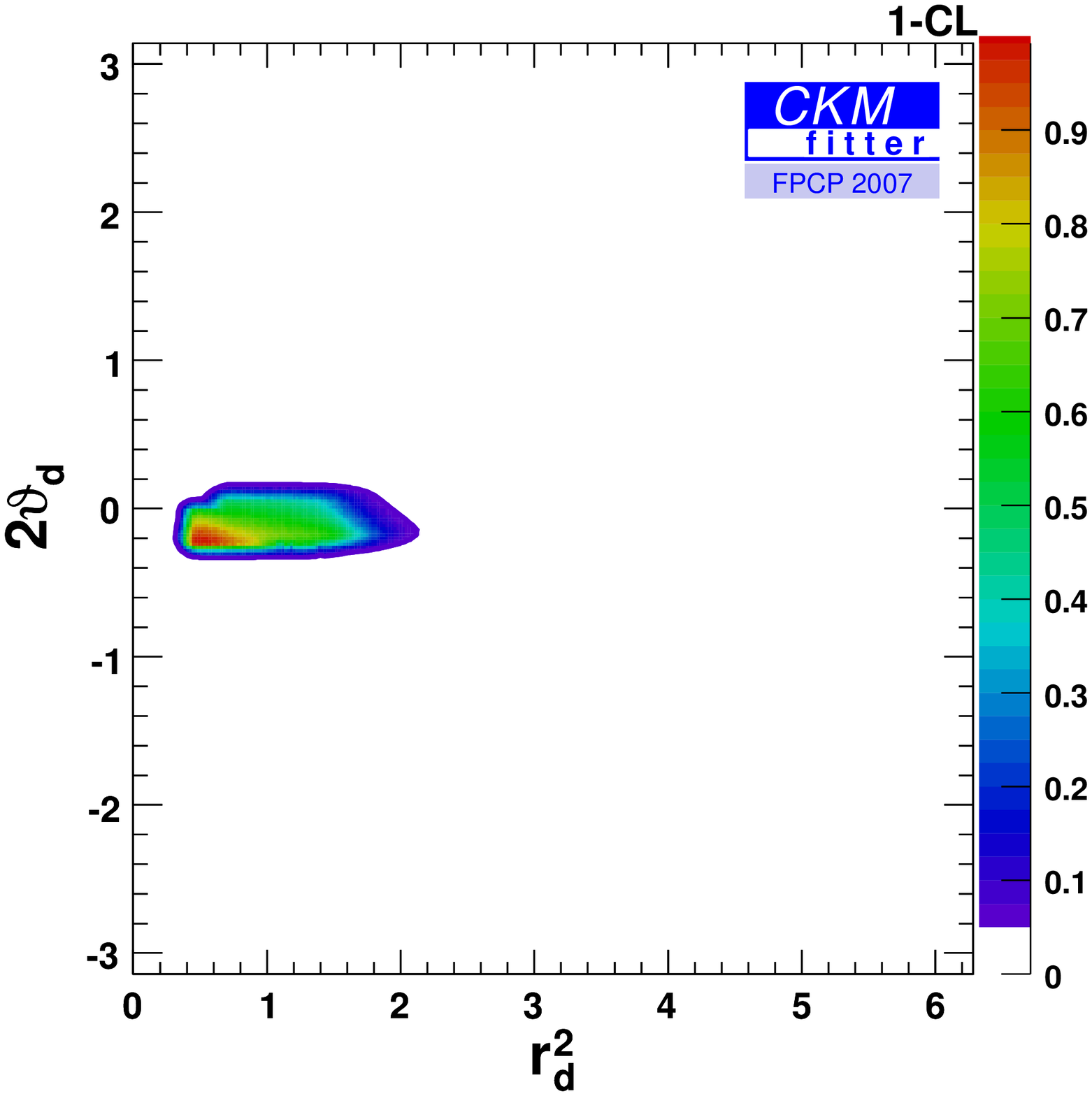}
\hspace{2em}
\includegraphics[width=2.85in]{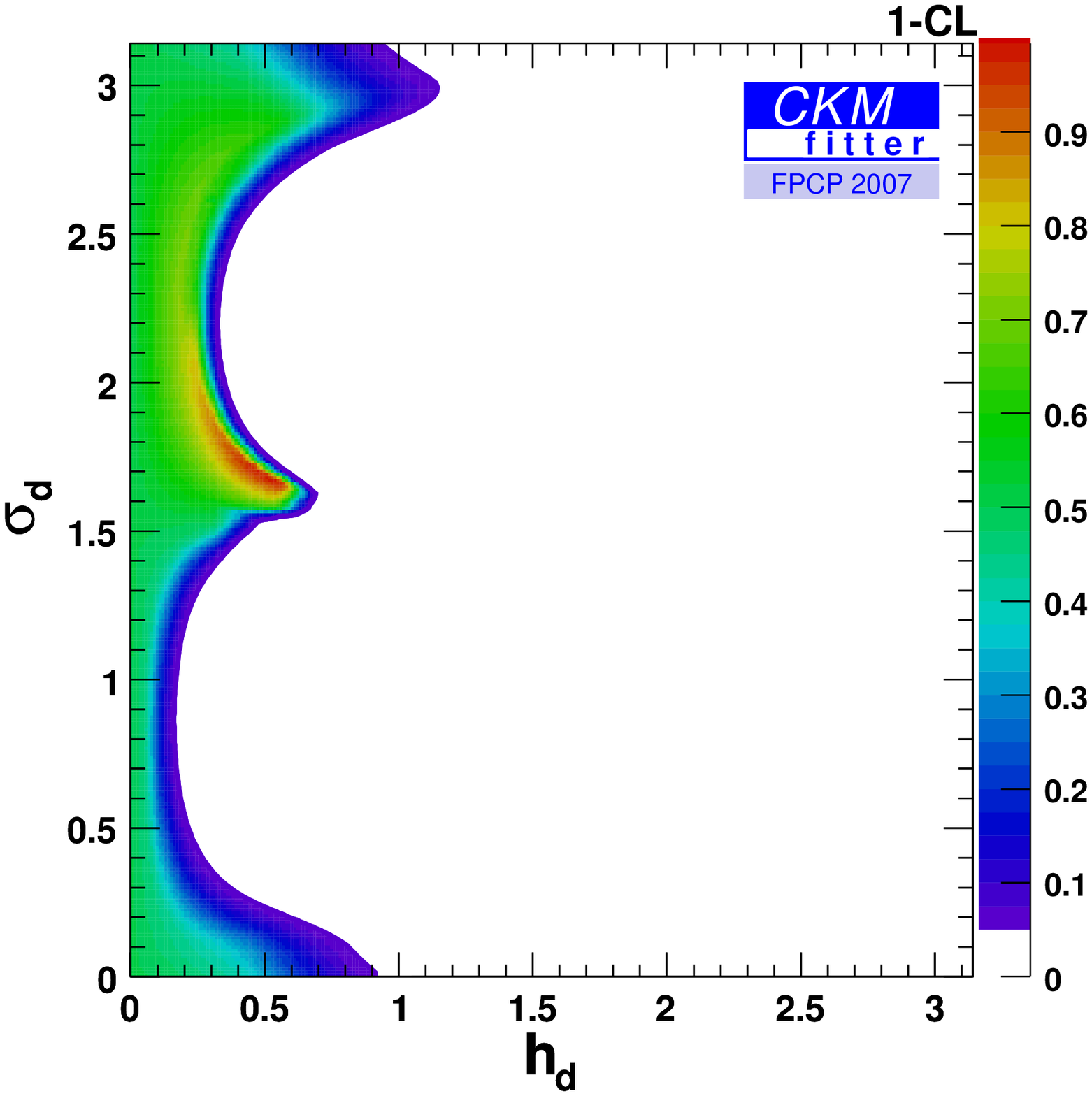}
\end{center}
\end{figure}

All that we need to do in order to establish whether the SM dominates
the observed CP violation, and to put an upper bound on the new
physics contribution to $B^0-\overline{B}{}^0$ mixing, is to project
the results of the fit performed in the previous subsection on the
$r_d^2-2\theta_d$ plane. If we find that $\theta_d\ll\beta$, then the 
SM dominance in the observed CP violation will be established. 
The constraints are shown in Fig.~\ref{fig:rdtd}(a). Indeed,
$\theta_d\ll\beta$.   

An alternative way to present the data is to use the $h_d,\sigma_d$
parametrization,
\beq
r_d^2e^{2i\theta_d}=1+h_d e^{i\sigma_d}.
\eeq
While the $r_d,\theta_d$ parameters give the relation between the full
mixing amplide and the SM one, and are convenient to apply to the
measurements, the $h_d,\sigma_d$ parameters give the relation between
the new physics and SM contributions, and are more convenient in
testing theoretical models:
\beq
h_de^{i\sigma_d}=\frac{M_{12}^{\rm NP}}{M_{12}^{\rm SM}}.
\eeq
The constraints in the $h_d-\sigma_d$ plane are shown in
Fig.~\ref{fig:rdtd}(b). We conclude that {\bf a new physics
contribution to the $B^0-\overline{B}^0$ mixing amplitude at a level
higher than about 30\% is now disfavored}.  

%%%%%%%%%%%
%%%%%%%%%
\section{Neutrino Anarchy versus Quark Hierarchy}
\label{sec:nu}
%%%%%%%%
A detailed presentation of the physics and the formalism of neutrino
flavor transitions is given in Appendix \ref{sec:nufl} for both vacuum
oscillations (\ref{sec:vac}) and the matter transitions
(\ref{sec:mat}). It follows Ref. \cite{Gonzalez-Garcia:2002dz}. 

{\bf Exercise 10:} {\it For atmospheric $\nu_\mu$'s with $E\sim1\ GeV$, the
flux coming from above has $P_{\mu\mu}(L\sim10\ {\rm km})\approx1$, while
the flux from below has $P_{\mu\mu}(L\sim10^4\ {\rm km})\approx0.5$. Assuming
that for the flux coming from below the oscillations are averaged
out, estimate $\Delta m^2$ and $\sin^22\theta$.}

{\bf Exercise 11:} {\it For solar $\nu_e$'s, the transition between matter
($\beta_{\rm MSW}>1$) and vacuum ($\beta_{\rm MSW}<\cos2\theta$)
flavor transitions occurs around $E\sim2\ MeV$. The transition
probability is measured to be roughly $P_{ee}\sim0.30$ for $\beta_{\rm
  MSW}>1$. Estimate $\Delta m^2$ and $\theta$ and predict
$P_{ee}$ for $\beta_{\rm MSW}\ll1$.} 

The derived ranges for the three mixing angles and two mass-squared
differences at $1\sigma$ are \cite{Gonzalez-Garcia:2007ib}:
\beqa\label{nupara}
\Delta m^2_{21}&=&(7.9\pm0.3)\times10^{-5}\ eV^2,\ \ \
|\Delta m^2_{32}|=(2.6\pm0.2)\times10^{-3}\ eV^2,\no\\
\sin^2\theta_{12}&=&0.31\pm0.02,\ \ \
\sin^2\theta_{23}=0.47\pm0.07,\ \ \
\sin^2\theta_{13}=0^{+0.008}_{-0.0}.
\eeqa
The $3\sigma$ range for the matrix elements of $U$ are the following
\cite{Gonzalez-Garcia:2007ib}: 
\beq\label{uthsi}
|U|=\pmatrix{0.79\to0.86&0.50\to0.61&0.00\to0.20\cr
  0.25\to0.53&0.47\to0.73&0.56\to0.79\cr
  0.21\to0.51&0.42\to0.69&0.61\to0.83\cr}.
\eeq

%%%%%%%%%%
\subsection{New physics}
The simplest and most straightforward lesson of the evidence for
neutrino masses is also the most striking one: there is new physics
beyond the Standard Model. This is the first experimental result that
is inconsistent with the SM.

Most likely, the new physics is related to the existence of $G_{\rm
  SM}$-singlet fermions at some high energy scale that induce, at low
energies, the effective terms of Eq. (\ref{Hnint}) through the seesaw
mechanism. The existence of heavy singlet fermions is predicted by
many extensions of the SM, especially by GUTs [beyond $SU(5)$] and
left-right-symmetric theories.

There are of course other possibilities. Neutrino masses can be
generated without introducing any new fermions beyond those of the SM.
Instead, the existence of a scalar $\Delta_L(1,3)_{+1}$, that is, an
$SU(2)_{\rm L}$-triplet, is required. The smallness of the neutrino
masses is related here to the smallness of the vacuum expectation
value $\langle\Delta_L^0\rangle$ (required also by the success of the
$\rho=1$ relation) and does not have a generic natural explanation.

In left-right-symmetric models, however, where the breaking of
$SU(2)_{\rm R}\times U(1)_{\rm B-L}\to U(1)_{\rm Y}$ is induced by the
VEV of an $SU(2)_{\rm R}$-triplet, $\Delta_R$, there must exist also
an $SU(2)_{\rm L}$-triplet scalar. Furthermore, the Higgs potential
leads to an order of magnitude relation between the various VEVs,
$\langle\Delta_L^0\rangle\langle\Delta_R^0\rangle\sim v^2$, and the
smallness of $\langle\Delta_L^0\rangle$ is correlated with the high
scale of $SU(2)_{\rm R}$ breaking. This situation can be thought of as
a seesaw of VEVs. In this model there are, however, also SM-singlet
fermions. The light neutrino masses arise from both the seesaw
mechanism (``type I'') and the triplet VEV (``type II'').

Neutrino masses could also be of the Dirac type. Here, again, singlet
fermions are introduced, but lepton number is imposed by hand. This
possibility is disfavored by theorists since it is likely that global
symmetries are violated by gravitational effects. Furthermore, the
lightness of the neutrinos (compared to charged fermions) is
unexplained.

Another possibility is that neutrino masses are generated by mixing
with singlet fermions but the mass scale of these fermions is not
high. Here again the lightness of neutrino masses remains a
puzzle. The best known example of such a scenario is the framework of
supersymmetry without $R$ parity.

Let us emphasize that the seesaw mechanism or, more generally, the
extension of the SM with non-renormalizable terms, is the simplest
explanation of neutrino masses. Models in which neutrino masses are
generated by new physics at low energy imply a much more dramatic
departure from the SM. Furthermore, the existence of seesaw masses is
an unavoidable prediction of various extensions of the SM. In
contrast, many (but not all) of the low energy mechanisms are
introduced for the specific purpose of generating neutrino masses.

%%%%%%%%%%
\subsection{The scale of new physics}
Eq. (\ref{Hnint}) gives a light neutrino mass matrix:
\beq\label{seesawmass}
(M_\nu)_{ij}=Z_{ij}^\nu\frac{v^2}{\Lambda_{\rm NP}}.
\eeq
It is straightforward to use the measured neutrino masses of
Eq. (\ref{nupara}) in combination with Eq. (\ref{seesawmass}) to
estimate the scale of new physics that is relevant to their
generation. In particular, if there is no quasi-degeneracy in the
neutrino masses, the heaviest of the active neutrino masses can be
estimated:
\beq\label{mthree}
m_h=m_3\sim\sqrt{\Delta m^2_{32}}\approx0.05\ eV.
\eeq
(In the case of inverted hierarchy, the implied scale is
$m_h=m_2\sim\sqrt{\Delta m^2_{32}}\approx0.05\ eV$.) It follows that
the scale in the nonrenormalizable terms (\ref{Hnint}) is given by
\beq\label{seesawlnp}
\Lambda_{\rm NP}\sim v^2/m_h\approx10^{15}\ GeV.
\eeq
We should clarify two points regarding Eq. (\ref{seesawlnp}):
\begin{enumerate}
\item There could be some level of degeneracy between the neutrino
  masses. In such a case, Eq. (\ref{mthree}) is modified into a lower
  bound on $m_3$ and, consequently, Eq. (\ref{seesawlnp}) becomes an
  upper bound on $\Lambda_{\rm NP}$.
\item It could be that the $Z_{ij}$ of Eq. (\ref{Hnint}) are much
  smaller than 1. In such a case, again, Eq. (\ref{seesawlnp}) becomes
  an upper bound on the scale of new physics.
\end{enumerate}

On the other hand, in models of approximate flavor symmetries, there
are relations between the structures of the charged lepton and
neutrino mass matrices that give, quite generically, $Z_{33}\gsim
m_\tau^2/v^2\sim10^{-4}$. We conclude that the likely range for
$\Lambda_{\rm NP}$ is given by
\beq\label{lnpssfl}
10^{11}\ GeV\lsim\Lambda_{\rm NP}\lsim10^{15}\ GeV.
\eeq

The estimates (\ref{seesawlnp}) and (\ref{lnpssfl}) are very
exciting. First, the upper bound on the scale of new physics is well
below the Planck scale. This means that there is new physics in Nature
which is intermediate between the two known scales, the Planck scale,
$m_{\rm Pl}\sim10^{19}\ GeV$, and the electroweak breaking scale,
$v\sim 10^2\ GeV$.

Second, the scale $\Lambda_{\rm NP}\sim10^{15}\ GeV$ is intriguingly
close to the scale of gauge coupling unification.

Third, the range (\ref{lnpssfl}) for the scale of lepton number
breaking is optimal for leptogenesis \cite{Fukugita:1986hr}. If
leptogenesis is generated by 
the decays of the lightest singlet neutrino $N_1$, and the masses of
the singlet neutrinos are hierarchical, $M_1/M_{2,3\ldots}\ll1$ , then
there is an upper bound on the CP asymmetry in $N_1$ decays
\cite{Davidson:2002qv}:
\beq
|\epsilon_{N_1}|\leq\frac{3}{16\pi}\frac{M_1(m_3-m_2)}{v^2}.
\eeq
Given that $Y_B^{\rm obs}\sim9\times10^{-11}$, and that
$Y_B\sim10^{-3}\eta\epsilon_{N_1}$, where $\eta\lsim1$ is a washout
factor, we must require $|\epsilon_{N_1}|\gsim10^{-7}$. Moreover, we
have $m_3-m_2\leq\sqrt{\Delta m^2_{32}}\sim0.05\ eV$ and therefore
obtain $M_1\gsim10^{9}\ GeV$. 

%%%%%%%%%%%%%%%%%%%%%
\subsection{The flavor puzzle}
In the absence of neutrino masses, there are 13 flavor parameters in
the SM:
\beqa\label{chafla}
y_t&\sim&1,\ \ y_c\sim10^{-2},\ \ y_u\sim10^{-5},\no\\
y_b&\sim&10^{-2},\ \ y_s\sim10^{-3},\ \ y_d\sim10^{-4},\no\\
y_\tau&\sim&10^{-2},\ \ y_\mu\sim10^{-3},\ \ y_e\sim10^{-6},\no\\
|V_{us}|&\sim&0.2,\ \ |V_{cb}|\sim0.04,\ \ |V_{ub}|\sim0.004,\ \
\sin\delta_{\rm KM}\sim1.
\eeqa
These flavor parameters are hierarchical (their magnitudes span six
orders of magnitude), and all but two or three (the top Yukawa, the CP
violating phase, and perhaps the Cabibbo angle) are small. The
unexplained smallness and hierarchy pose the SM {\it flavor puzzle}.
Its solution may direct us to physics beyond the Standard Model.

Several mechanisms have been proposed in response to this puzzle. For
example, approximate horizontal symmetries, broken by a small
parameter, can lead to selection rules that explain the hierarchy of
the Yukawa couplings.

In the extension of the SM with three active neutrinos that have
Majorana masses, there are nine new flavor parameters in addition to
those of Eq. (\ref{chafla}). These are three neutrino
masses, three lepton mixing angles, and three phases in the mixing
matrix. Of the nine new parameters, four have been measured: two
mass-squared differences and two mixing angles [see Eq.
(\ref{nupara})]. This adds significantly to the input data on flavor
physics and provides an opportunity to test and refine flavor models.

If neutrino masses arise from effective terms of the form of
Eq. (\ref{Hnint}), then the overall scale of neutrino masses is
related to the scale $\Lambda_{\rm NP}$ and, in most cases, does not
tell us anything about flavor physics. More significant information
for flavors models can be written in terms of three dimensionless
parameters whose values can be read from Eq. (\ref{nupara}), that is
$\sin\theta_{12}$, $\sin\theta_{23}$ and 
\beq\label{nuflpa}
\Delta m^2_{21}/|\Delta m^2_{32}|=0.030\pm0.003.
\eeq
In addition, the upper bound on $\sin\theta_{13}$ often plays a
significant role in flavor model building. 

There are several features in the numerical estimates
(\ref{nupara},\ref{nuflpa}) that have drawn much attention and have
driven numerous investigations:

(i) Large mixing and strong hierarchy: The mixing angle that is
relevant to the $2-3$ sector is large, $\sin\theta_{23}\sim0.7$. On
the other hand, if there is no quasi-degeneracy in the neutrino
masses, the corresponding mass ratio is small, $m_2/m_3\sim0.17$. It
is difficult to explain in a natural way a situation where there is an
${\cal O}(1)$ mixing but the corresponding masses are hierarchical.

(ii) Two large and one small mixing angles: The mixing angles relevant
to the $2-3$ sector ($\sin\theta_{23}\sim0.7$) and $1-2$ sector
($\sin\theta_{12}\sim0.55$) are large, yet the $1-3$ mixing angle is
small ($\sin\theta_{13}\lsim 0.20$). Such a situation is, again,
difficult -- though not impossible -- to explain from approximate
symmetries. An example of a symmetry that does predict such a pattern
is that of $L_e-L_\mu-L_\tau$. This symmetry predicts, however,
$\theta_{12}\simeq\pi/4$, which is experimentally excluded.

(iii) Maximal mixing: The value of $\theta_{23}$ is intriguingly close
to maximal mixing ($\sin^22\theta_{23}=1$). It is interesting to
understand whether a symmetry could explain this special value.

(iv) Tribimaximal mixing: The mixing matrix (\ref{uthsi}) has a
structure that is consistent with the following unitary matrix
\cite{Harrison:2002er}: 
\beq
U=\pmatrix{\sqrt{\frac23}&\sqrt{\frac13}&0\cr
  -\sqrt{\frac16}&\sqrt{\frac13}&\sqrt{\frac12}\cr
\sqrt{\frac16}&-\sqrt{\frac13}&\sqrt{\frac12}\cr}.
\eeq
It is interesting to understand whether a symmetry could explain this
special structure.

All four features enumerated above are difficult to explain in a large
class of flavor models that do very well in explaining the flavor
features of the quark sector. In particular, models with Abelian
horizontal symmetries (Froggatt-Nielsen type \cite{Froggatt:1978nt})
predict that, in general, $|V_{ub}|\sim|V_{us}V_{cb}|$, $|V_{ij}|\gsim
m_i/m_j$ ($i<j$) and $V\sim{\bf 1}$
\cite{Leurer:1992wg,Leurer:1993gy}. All of these are successful
predictions. At the same time, however, these models predict
\cite{Grossman:1995hk} that for the neutrinos, in general,
$|U_{ij}|^2\sim m_i/m_j$ and $|U_{e3}|\sim|U_{e2}U_{\mu3}|$, in
contradiction to, respectively, points (i) and (ii) above (and there
is no way to make $\theta_{23}$ parametrically close to $\pi/4$). On
the other hand, there exist very specific models where these features
are related to a symmetry.

It is possible, however, that the above interpretation of the results
is wrong. Indeed, the data can be interpreted in a very different
way:

(v) No small parameters. The two measured mixing angles are larger
than any of the quark mixing angles. Indeed, they are both of order
one. The measured mass ratio, $m_2/m_3\gsim0.16$ is larger than any of
the quark and charged lepton mass ratios, and could be interpreted as
an ${\cal O}(1)$ parameter (namely, it is accidentally small, without
any parametric suppression). If this is the correct way of reading the
data, the measured neutrino parameters may actually reflect the
absence of any hierarchical structure in the neutrino mass matrices
\cite{Hall:1999sn}. The possibility that there is no structure --
neither hierarchy, nor degeneracy -- in the neutrino sector has been
called ``neutrino mass anarchy''. An important test of this idea will
be provided by the measurement of $|U_{e3}|$. If indeed the entries in
$M_\nu$ have random values of the same order, all three mixing angles
are expected to be of order one. If experiments measure
$|U_{e3}|\sim0.1$, that is, close to the present bound, it can be
argued that its smallness is accidental. The stronger the upper bound
on this angle becomes, the more difficult it will be to maintain this
view.

Neutrino mass anarchy can be accommodated within models of Abelian
flavor symmetries, if the three lepton doublets carry the same
charge. Indeed, consider a supersymmetric model with a $U(1)_H$
symmetry that is broken by a single small spurion $\epsilon_H$ of
charge $-1$. Let us assume that the three fermion generations
contained in the $10$-representation of $SU(5)$ carry charges
$(2,1,0)$, while the three $\bar5$-representations carry charges
$(0,0,0)$. (The Higgs fields carry no $H$ charges.) Such a model
predicts $\epsilon_H^2$ hierarchy in the up sector, $\epsilon_H$
hierarchy in the down and charged lepton sectors, and anarchy in the
neutrino sector.

{\bf Exercise 12:} {\it The selection rule for this model is that a term in the
superpotential that carries $H$ charge $n\geq0$ is suppressed  by
$\epsilon_H^n$. Find the parametric suppression of the various entries
in $M_u,M_d,M_\ell$ and $M_\nu$. Find the parametric suppression of
the mixing angles.}

It would be nice if the features of quark mass hierarchy and neutrino
mass anarchy can be traced back to some fundamental principle or to a
stringy origin (see, for example, \cite{Antebi:2005hr}).

%%%%%%%%%%%%%%%%%%%%%%%%%%%%
\section{Conclusions}
\label{sec:con}
We have described four topics in flavor physics, each demonstrating a
different point of interest:

(i) The upper bound on $\Delta m_D$ shows that alignment cannot be the
only flavor mechanism that suppresses the supersymmetric flavor
changing contributions. It demonstrates how flavor physics at the GeV
scale probes new physics at the TeV scale.

(ii) The measurement of $S_{\psi K}$ provides a precision test of the
Kobayashi-Maskawa mechanism of CP violation. It strengthens the
evidence that this is the dominant source of CP violation in flavor
changing processes.

(iii) The LHC may discover new particles that couple to the standard
model fermions. If that happens, we will be able to use the new
physics for better understanding of the flavor puzzle, and the flavor
physics for better understanding of the new physics.

(iv) The measurements of neutrino flavor parameters -- mass-squared
differences and mixing angles -- have tested models that aim to
explain the hierarchy in the quark sector, and have added novel
aspects to the question of whether the flavor structure has a
symmetry-related explanation.

The huge progress in flavor physics in recent years has provided
answers to many questions. At the same time, new questions arise. We
look forward to the LHC era for more answers and more questions.

%%%%%%%%%%%%%%%%%%%%%%%%%%%%%
\section*{Acknowledgments}
I thank Michael Dine and Nathan Seiberg for their hospitality at the
IAS, and Gian Giudice, Michelangelo Mangano and David Jacobs for their 
hospitality at CERN. 
I am grateful to Heiko Lacker, Stephane T'Jampens and the CKMfitter
group for their help. I thank Gil Paz for comments on the manuscript.
The research of Y.N. is supported by the Israel Science Foundation,
the United States-Israel Binational Science Foundation (BSF),
Jerusalem, Israel, the German-Israeli foundation for scientific
research and development (GIF), and the Minerva Foundation.

%%%%%%%%%%%%%%%%%%%%%%%%%%%%%
\appendix
%%%%%%%%%%%%%%
%%%%%%%%%%%%%%
\section{The CKM matrix}
\label{app:ckm}
The CKM matrix $V$ is a $3\times3$ unitary matrix. Its form, however, 
is not unique:

$(i)$ There is freedom in defining $V$ in that we can permute between
the various generations. This freedom is fixed by ordering the up quarks and 
the down quarks by their masses, {\it i.e.} $(u_1,u_2,u_3)\to(u,c,t)$ and 
$(d_1,d_2,d_3)\to(d,s,b)$. The elements of $V$ are written as follows:
\beq\label{defVij}
V=\pmatrix{V_{ud}&V_{us}&V_{ub}\cr
V_{cd}&V_{cs}&V_{cb}\cr V_{td}&V_{ts}&V_{tb}\cr}.
\eeq

$(ii)$ There is further freedom in the phase structure of $V$. This
means that the number of physical parameters in $V$ is smaller than
the number of parameters in a general unitary $3\times3$ matrix which
is nine (three real angles and six phases). Let us define $P_q$
($q=u,d$) to be diagonal unitary (phase) matrices. Then, if instead of
using $V_{qL}$ and $V_{qR}$ for the rotation (\ref{diagMq}) to the
mass basis we use $\tilde V_{qL}$ and $\tilde V_{qR}$, defined by
$\tilde V_{qL}=P_q V_{qL}$ and $\tilde V_{qR}=P_q V_{qR}$, we still
maintain a legitimate mass basis since $M_q^{\rm diag}$ remains
unchanged by such transformations. However, $V$ does change:
\beq\label{eqphase}
V\to P_u V P_d^*.
\eeq 
This freedom is fixed by demanding that $V$ has the minimal number of
phases. In the three generation case $V$ has a single phase. (There 
are five phase differences between the elements of $P_u$ and $P_d$ and, 
therefore, five of the six phases in the CKM matrix can be removed.) This is 
the Kobayashi-Maskawa phase $\delta_{\rm KM}$ which is the single source of 
CP violation in the quark sector of the Standard Model \cite{Kobayashi:1973fv}. 

The fact that $V$ is unitary and depends on only four independent
physical parameters can be made manifest by choosing a specific
parametrization. The standard choice is \cite{Chau:1984fp}
\beq\label{stapar}
V=\pmatrix{c_{12}c_{13}&s_{12}c_{13}&
s_{13}e^{-i\delta}\cr 
-s_{12}c_{23}-c_{12}s_{23}s_{13}e^{i\delta}&
c_{12}c_{23}-s_{12}s_{23}s_{13}e^{i\delta}&s_{23}c_{13}\cr
s_{12}s_{23}-c_{12}c_{23}s_{13}e^{i\delta}&
-c_{12}s_{23}-s_{12}c_{23}s_{13}e^{i\delta}&c_{23}c_{13}\cr},
\eeq
where $c_{ij}\equiv\cos\theta_{ij}$ and $s_{ij}\equiv\sin\theta_{ij}$.
The $\theta_{ij}$'s are the three real mixing parameters while
$\delta$ is the Kobayashi-Maskawa phase. It is known experimentally
that $s_{13}\ll s_{23}\ll s_{12}\ll1$. It is convenient to choose an
approximate expression where this hierarchy is manifest. This is the
Wolfenstein parametrization, where the four mixing parameters are
$(\lambda,A,\rho,\eta)$ with $\lambda=|V_{us}|=0.23$ playing the role
of an expansion parameter and $\eta$ representing the CP violating
phase \cite{Wolfenstein:1983yz,Buras:1994ec}:
\beq\label{wolpar}
V=\pmatrix{
1-\frac12\lambda^2-\frac18\lambda^4 & \lambda &
A\lambda^3(\rho-i\eta)\cr
-\lambda +\frac12A^2\lambda^5[1-2(\rho+i\eta)] &
1-\frac12\lambda^2-\frac18\lambda^4(1+4A^2) & A\lambda^2 \cr
A\lambda^3[1-(1-\frac12\lambda^2)(\rho+i\eta)]&
-A\lambda^2+\frac12A\lambda^4[1-2(\rho+i\eta)]
& 1-\frac12A^2\lambda^4 \cr}\; .
\eeq

\begin{figure}[tb]
  \centering
  {\includegraphics[width=0.65\textwidth]{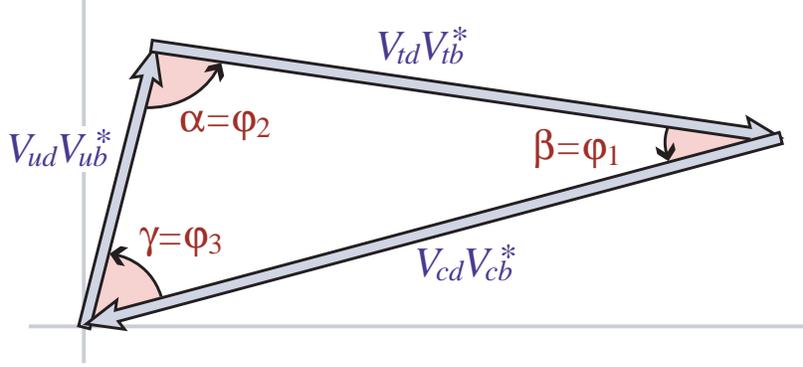}}
  \caption{Graphical representation of the unitarity constraint
  $V_{ud}V_{ub}^*+V_{cd}V_{cb}^*+V_{td}V_{tb}^*=0$ as a triangle in
  the complex plane.}
  \label{fg:tri}
\end{figure}

A very useful concept is that of the {\it unitarity triangles}. The
unitarity of the CKM matrix leads to various relations among the
matrix elements, {\it e.g.}
\beqa\label{Unitds}
V_{ud}V_{us}^*+V_{cd}V_{cs}^*+V_{td}V_{ts}^*=0,\\
\label{Unitsb}
V_{us}V_{ub}^*+V_{cs}V_{cb}^*+V_{ts}V_{tb}^*=0,\\
\label{Unitdb}
V_{ud}V_{ub}^*+V_{cd}V_{cb}^*+V_{td}V_{tb}^*=0.
\eeqa
Each of these three relations requires 
the sum of three complex quantities to vanish and so can be geometrically
represented in the complex plane as a triangle. These are
``the unitarity triangles", though the term ``unitarity triangle"
is usually reserved for the relation (\ref{Unitdb}) only. The
unitarity triangle related to Eq. (\ref{Unitdb}) is depicted in
Fig. \ref{fg:tri}. 

The rescaled unitarity triangle  is derived from (\ref{Unitdb})
by (a) choosing a phase convention such that $(V_{cd}V_{cb}^*)$
is real, and (b) dividing the lengths of all sides by $|V_{cd}V_{cb}^*|$.
Step (a) aligns one side of the triangle with the real axis, and
step (b) makes the length of this side 1. The form of the triangle
is unchanged. Two vertices of the rescaled unitarity triangle are
thus fixed at (0,0) and (1,0). The coordinates of the remaining
vertex correspond to the Wolfenstein parameters $(\rho,\eta)$.
The area of the rescaled unitarity triangle is $|\eta|/2$.

Depicting the rescaled unitarity triangle in the
$(\rho,\eta)$ plane, the lengths of the two complex sides are
\beq\label{RbRt}
R_u\equiv\left|{V_{ud}V_{ub}\over V_{cd}V_{cb}}\right|
=\sqrt{\rho^2+\eta^2},\ \ \
R_t\equiv\left|{V_{td}V_{tb}\over V_{cd}V_{cb}}\right|
=\sqrt{(1-\rho)^2+\eta^2}.
\eeq
The three angles of the unitarity triangle are defined as follows 
\cite{Dib:1989uz,Rosner:1988nx}:
\beq\label{abcangles}
\alpha\equiv\arg\left[-{V_{td}V_{tb}^*\over V_{ud}V_{ub}^*}\right],\ \ \
\beta\equiv\arg\left[-{V_{cd}V_{cb}^*\over V_{td}V_{tb}^*}\right],\ \ \
\gamma\equiv\arg\left[-{V_{ud}V_{ub}^*\over V_{cd}V_{cb}^*}\right].
\eeq
They are physical quantities and can be independently measured by CP
asymmetries in $B$ decays. It is also useful to define the two 
small angles of the unitarity triangles (\ref{Unitsb},\ref{Unitds}):
\beq\label{bbangles}
\beta_s\equiv\arg\left[-{V_{ts}V_{tb}^*\over V_{cs}V_{cb}^*}\right],\ \ \
\beta_K\equiv\arg\left[-{V_{cs}V_{cd}^*\over V_{us}V_{ud}^*}\right].
\eeq

The $\lambda$ and $A$ parameters are very well determined at present,
see Eq. (\ref{lamaexp}). The main effort in CKM measurements is thus
aimed at improving our knowledge of $\rho$ and $\eta$:
\beq
\rho=0.14^{+0.04}_{-0.02},\ \ \ \eta=0.35\pm0.02.
\eeq
The present status of our knowledge is best seen in a plot of the
various constraints and the final allowed region in the $\rho-\eta$
plane. This is shown in Fig. \ref{fg:UT}.

%%%%%%%%%%%%%%%%%%%%%%
%%%%%%%%%%%%%%%%%%%%%%
\section{Supersymmetric Contributions to Neutral Meson Mixing}
\label{app:susyd}
We consider the squark-gluino box diagram contribution to
$D^0-\overline{D}^0$ mixing amplitude that is proportional to
$K_{2i}^u K^{u*}_{1i}K_{2j}^u K^{u*}_{1j}$, where $K^u$ is the
mixing matrix of the gluino couplings to left-handed up quarks and
their up squark partners. (In the language of the mass insertion
approximation, we calculate here the contribution that is $\propto
[(\delta^u_{LL})_{12}]^2$.) We work in the mass basis for both quarks
and squarks.

The contribution is given by
\beq\label{motsusy}
M_{12}^D=-i\frac{4\pi^2}{27}\alpha_s^2m_Df_D^2B_D\eta_{\rm QCD}
\sum_{i,j}(K_{2i}^uK_{1i}^{u*}K_{2j}^uK_{1j}^{u*})(11\tilde
I_{4ij}+4\tilde m_g^2I_{4ij}).
\eeq
where
\beqa
\tilde I_{4ij}&\equiv&\int\frac{d^4p}{(2\pi)^4}\frac{p^2}{(p^2-\tilde
  m_g^2)^2(p^2-\tilde m_i^2)(p^2-\tilde m_j^2)}\no\\
&=&\frac{i}{(4\pi)^2}\left[\frac{\tilde m_g^2}
  {(\tilde m_i^2-\tilde m_g^2)(\tilde m_j^2-\tilde m_g^2)}\right.\no\\
   && +\left.\frac{\tilde m_i^4}
  {(\tilde m_i^2-\tilde m_j^2)(\tilde m_i^2-\tilde
    m_g^2)^2}\ln\frac{\tilde m_i^2}{\tilde m_g^2}
  +\frac{\tilde m_j^4}
  {(\tilde m_j^2-\tilde m_i^2)(\tilde m_j^2-\tilde
    m_g^2)^2}\ln\frac{\tilde m_j^2}{\tilde m_g^2}\right],
\eeqa
\beqa
I_{4ij}&\equiv&\int\frac{d^4p}{(2\pi)^4}\frac{1}{(p^2-\tilde
  m_g^2)^2(p^2-\tilde m_i^2)(p^2-\tilde m_j^2)}\no\\
&=&\frac{i}{(4\pi)^2}\left[\frac{1}
  {(\tilde m_i^2-\tilde m_g^2)(\tilde m_j^2-\tilde m_g^2)}\right.\no\\
   && +\left.\frac{\tilde m_i^2}
  {(\tilde m_i^2-\tilde m_j^2)(\tilde m_i^2-\tilde
    m_g^2)^2}\ln\frac{\tilde m_i^2}{\tilde m_g^2}
  +\frac{\tilde m_j^2}
  {(\tilde m_j^2-\tilde m_i^2)(\tilde m_j^2-\tilde
    m_g^2)^2}\ln\frac{\tilde m_j^2}{\tilde m_g^2}\right].
\eeqa

We now follow the discussion in refs. \cite{Raz:2002zx,Nir:2002ah}. 
To see the consequences of the super-GIM mechanism, let us expand the
expression for the box integral around some value $\tilde m^2_q$ for
the squark masses-squared:
\beqa
I_4(\tilde m_g^2,\tilde m_i^2,\tilde m_j^2)&=&
I_4(\tilde m_g^2,\tilde m_q^2+\delta\tilde m_i^2,\tilde
m_q^2+\delta\tilde m_j^2)\no\\
&=&I_4(\tilde m_g^2,\tilde m_q^2,\tilde m_q^2)
+(\delta\tilde m_i^2+\delta\tilde m_j^2)I_5(\tilde m_g^2,\tilde
m_q^2,\tilde m_q^2,\tilde m_q^2)\no\\
&+&\frac12\left[(\delta\tilde m_i^2)^2+(\delta\tilde
  m_j^2)^2+2(\delta\tilde m_i^2)(\delta\tilde m_j^2)\right]I_6(\tilde m_g^2,\tilde
m_q^2,\tilde m_q^2,\tilde m_q^2,\tilde m_q^2)+\cdots,
\eeqa
where
\beq
I_n(\tilde m_g^2,\tilde m_q^2,\ldots,\tilde
m_q^2)\equiv\int\frac{d^4p}{(2\pi)^4}\frac{1}{(p^2-\tilde
  m_g^2)^2(p^2-\tilde m_q^2)^{n-2}},
\eeq
and similarly for $\tilde I_{4ij}$. Note that $I_n\propto(\tilde
m_q^2)^{n-2}$ and $\tilde I_n\propto(\tilde m_q^2)^{n-3}$. Thus, using
$x\equiv\tilde m_g^2/\tilde m_q^2$, it is customary to define
\beq
I_n\equiv\frac{i}{(4\pi)^2(\tilde m_q^2)^{n-2}}f_n(x),\ \ \ \
\tilde I_n\equiv\frac{i}{(4\pi)^2(\tilde m_q^2)^{n-3}}\tilde f_n(x).
\eeq
The unitarity of the mixing matrix implies that
\beq
\sum_i (K_{2i}^uK_{1i}^{u*}K_{2j}^uK_{1j}^{u*})=
\sum_j (K_{2i}^uK_{1i}^{u*}K_{2j}^uK_{1j}^{u*})=0.
\eeq
We learn that the terms that are proportional $f_4,\tilde f_4,f_5$ and
$\tilde f_5$ vanish in their contribution to $M_{12}$. When
$\delta\tilde m_i^2\ll\tilde m_q^2$ for all $i$, the
leading contributions to $M_{12}$ come from $f_6$ and $\tilde f_6$. We
learn that for quasi-degenerate squarks, the leading contribution is
quadratic in the small mass-squared difference. The functions $f_6(x)$
and $\tilde f_6(x)$ are given by
\beqa
f_6(x)&=&\frac{6(1+3x)\ln x+x^3-9x^2-9x+17}{6(1-x)^5},\no\\
\tilde f_6(x)&=&\frac{6x(1+x)\ln x-x^3-9x^2+9x+1}{3(1-x)^5}.
\eeqa
For example, with $x=1$, $f_6(1)=-1/20$ and $\tilde f_6=+1/30$;
with $x=2.33$, $f_6(2.33)=-0.015$ and $\tilde f_6=+0.013$.

To further simplify things, let us consider a two generation
case. Then
\beqa
M_{12}^D&\propto& 2(K_{21}^uK_{11}^{u*})^2(\delta\tilde
m_1^2)^2+2(K_{22}^uK_{12}^{u*})^2(\delta\tilde
m_2^2)^2+(K_{21}^uK_{11}^{u*}K_{22}^uK_{12}^{u*})(\delta\tilde
m_1^2+\delta\tilde m_2^2)^2\no\\
&=&(K^u_{21}K_{11}^{u*})^2(\tilde m_2^2-\tilde m_1^2)^2.
\eeqa
We thus rewrite Eq.~(\ref{motsusy}) for the case of quasi-degenerate squarks:
\beq\label{motsusyd}
M_{12}^D=\frac{\alpha_s^2m_Df_D^2B_D\eta_{\rm QCD}}{108\tilde m_q^2}
[11\tilde f_6(x)+4xf_6(x)]\frac{(\Delta\tilde m^2_{21})^2}{\tilde m_q^4}
(K_{21}^uK_{11}^{u*})^2.
\eeq
For example, for $x=1$, $11\tilde f_6(x)+4xf_6(x)=+0.17$.
For $x=2.33$, $11\tilde f_6(x)+4xf_6(x)=+0.003$.

%%%%%%%%%%%%%%%%%%%%%%%%%%%
%%%%%%%%%%%%%%%%%%%%%%%%%%
\section{CP violation in neutral $B$ decays to final CP eigenstates}
\label{sec:formalism}
We define decay amplitudes of $B$ (which could be charged or neutral)
and its CP conjugate $\Bbar$ to a multi-particle final state $f$ and its
CP conjugate $\fb$ as
\beq\label{decamp}
A_{\f}=\langle \f|{\cal H}|B\rangle\quad , \quad
\overline{A}_{\f}=\langle \f|{\cal H}|\Bbar\rangle\quad , \quad
A_{\fb}=\langle \fb|{\cal H}|B\rangle\quad , \quad
\overline{A}_{\fb}=\langle \fb|{\cal H}|\Bbar\rangle\; ,
\eeq
where ${\cal H}$ is the Hamiltonian governing
weak interactions.  The action of CP on these states introduces
phases $\xi_B$ and $\xi_f$ according to
\beqa\label{eq:phaseconv}
\CP|B\rangle &=& e^{+i\xi_{B}}\,|\Bbar\rangle \quad , \quad
\CP|\f\rangle = e^{+i\xi_{\f}}\,|\fb\rangle \; ,\no\\
\CP|\Bbar\rangle& =& e^{-i\xi_{B}}\,|B\rangle \quad , \quad
\CP|\fb\rangle = e^{-i\xi_{\f}}\,|\f\rangle \ ,
\eeqa
so that $(\CP)^2=1$. The phases $\xi_B$ and $\xi_f$ are arbitrary and
unphysical because of the flavor symmetry of the strong
interaction. If CP is conserved by the dynamics, $[\CP,{\cal H}] =
0$, then $A_f$ and $\overline{A}_{\fb}$ have the same magnitude and an
arbitrary unphysical relative phase 
\beq\label{spupha}
\overline{A}_{\fb} = e^{i(\xi_{\f}-\xi_{B})}\, A_f\; .
\eeq

A state that is initially a superposition of $\Bz$ and $\Bzb$, say
\beq
|\psi(0)\rangle = a(0)|\Bz\rangle+b(0)|\Bzb\rangle \; ,
\eeq
will evolve in time acquiring components that describe all possible
decay final states $\{f_1,f_2,\ldots\}$, that is,
\beq
|\psi(t)\rangle =
a(t)|\Bz\rangle+b(t)|\Bzb\rangle+c_1(t)|f_1\rangle+c_2(t)|f_2\rangle+\cdots
\; .
\eeq 
If we are interested in computing only the values of $a(t)$ and $b(t)$
(and not the values of all $c_i(t)$), and if the times $t$ in which we
are interested are much larger than the typical strong interaction
scale, then we can use a much simplified
formalism~\cite{Weisskopf:au}. The simplified time evolution is
determined by a $2\times 2$ effective Hamiltonian $\Heff$ that is
not Hermitian, since otherwise the mesons would only oscillate and not
decay. Any complex matrix, such as $\Heff$, can be written in terms of
Hermitian matrices $\Meff$ and $\Geff$ as
\beq
\Heff = \Meff - \frac{i}{2}\,\Geff \; .
\eeq
$\Meff$ and $\Geff$ are associated with
$(\Bz,\Bzb)\leftrightarrow(\Bz,\Bzb)$ transitions via off-shell
(dispersive) and on-shell (absorptive) intermediate states, respectively.
Diagonal elements of $\Meff$ and $\Geff$ are associated with the
flavor-conserving transitions $\Bz\to\Bz$ and $\Bzb\to\Bzb$ while
off-diagonal elements are associated with flavor-changing transitions
$\Bz\leftrightarrow\Bzb$.

The eigenvectors of $\Heff$ have well defined masses and decay
widths. We introduce complex parameters $p_{L,H}$ and $q_{L,H}$ to
specify the components of the strong interaction eigenstates, $\Bz$ and
$\Bzb$, in the light ($B_L$) and heavy ($B_H$) mass eigenstates:
\beq\label{defpq}
|B_{L,H}\rangle=p_{L,H}|\Bz\rangle\pm q_{L,H}|\Bzb\rangle
\eeq
with the normalization $|p_{L,H}|^2+|q_{L,H}|^2=1$. If either CP or
CPT is a symmetry of $\Heff$ (independently of whether T is conserved or
violated) then $\Meff_{11} = \Meff_{22}$ and $\Geff_{11}=
\Geff_{22}$, and solving the eigenvalue problem for $\Heff$ yields $p_L
= p_H \equiv p$ and $q_L = q_H \equiv q$ with
\beq
\left(\frac{q}{p}\right)^2=\frac{\Meff_{12}^\ast -
    (i/2)\Geff_{12}^\ast}{\Meff_{12}-(i/2)\Geff_{12}}\; .
\eeq
From now on we assume that CPT is conserved.
If either CP or T is a symmetry of $\Heff$ (independently of whether
CPT is conserved or violated), then $\Meff_{12}$ and $\Geff_{12}$ are
relatively real, leading to
\beq
\left(\frac{q}{p}\right)^2 = e^{2i\xi_B} \quad \Rightarrow \quad
\left|\frac{q}{p}\right| = 1 \; ,
\eeq
where $\xi_B$ is the arbitrary unphysical phase introduced in
Eq.~(\ref{eq:phaseconv}).

The real and imaginary parts of the eigenvalues of $\Heff$
corresponding to $|B_{L,H}\rangle$ represent their masses and
decay-widths, respectively. The mass difference $\Delta m_B$ and the
width difference $\Delta\Gamma_B$ are defined as follows:
\beq\label{DelmG}
\Delta m_B\equiv M_H-M_L,\ \ \ \Delta\Gamma_B\equiv\Gamma_H-\Gamma_L.
\eeq
Note that here $\Delta m_B$ is positive by definition, while the sign of
$\Delta\Gamma_B$ is to be experimentally determined. 
The average mass and width are given by
\beq\label{aveMG}
m_B\equiv{M_H+M_L\over2},\ \ \ \Gamma_B\equiv{\Gamma_H+\Gamma_L\over2}.
\eeq
It is useful to define dimensionless ratios $x$ and $y$:
\beq\label{defxy}
x\equiv{\Delta m_B\over\Gamma_B},\ \ \ y\equiv{\Delta\Gamma_B\over2\Gamma_B}.
\eeq
Solving the eigenvalue equation gives
\beq\label{eveq}
(\Delta m_B)^2-{1\over4}(\Delta\Gamma_B)^2=(4|M_{12}|^2-|\Gamma_{12}|^2),\ \ \ \ 
\Delta m_B\Delta\Gamma_B=4\re{M_{12}\Gamma_{12}^*}.
\eeq

All CP-violating observables in $B$ and $\Bbar$ decays to final states $f$
and $\fb$ can be expressed in terms of phase-convention-independent
combinations of $A_f$, $\overline{A}_f$, $A_{\overline{f}}$ and
$\overline{A}_{\overline{f}}$, together with, for neutral-meson decays
only, $q/p$. CP violation in charged-meson decays depends only on the
combination $|\overline{A}_{\fb}/A_f|$, while CP violation in
neutral-meson decays is complicated by $\Bz\leftrightarrow\Bzb$
oscillations and depends, additionally, on $|q/p|$ and on $\lambda_f
\equiv (q/p)(\overline{A}_f/A_f)$.

For neutral $D$, $B$, and $B_s$ mesons, $\Delta\Gamma/\Gamma\ll1$ and
so both mass eigenstates must be considered in their evolution. We
denote the state of an initially pure $|\Bz\rangle$ or $|\Bzb\rangle$
after an elapsed proper time $t$ as $|\Bz_{\mathrm{phys}}(t)\rangle$
or $|\Bzb_{\mathrm{phys}}(t)\rangle$, respectively. Using the
effective Hamiltonian approximation, we obtain
\beqa\label{defphys}
|\Bz_{\rm phys}(t)\rangle&=&g_+(t)\,|\Bz\rangle
- \frac qp\ g_-(t)|\Bzb\rangle,\no\\
|\Bzb_{\rm phys}(t)\rangle&=&g_+(t)\,|\Bzb\rangle
- \frac pq\ g_-(t)|\Bz\rangle \; ,
\eeqa
where 
\beq
g_\pm(t) \equiv \frac12\left(e^{-im_Ht-\frac12\Gamma_Ht}\pm
  e^{-im_Lt-\frac12\Gamma_Lt}\right).
\eeq

One obtains the following time-dependent decay rates:
\beqa
\frac{d\Gamma[\Bz_{\rm phys}(t)\to f]/dt}{e^{-\Gamma t}{\cal N}_f}&=&
\left(|A_f|^2+|(q/p)\overline{A}_f|^2\right)\cosh(y\Gamma t)
  +\left(|A_f|^2-|(q/p)\overline{A}_f|^2\right)\cos(x\Gamma t)\no\\
&+&2\,\re{(q/p)A_f^\ast \overline{A}_f}\sinh(y\Gamma t)
-2\,\im{(q/p)A_f^\ast \overline{A}_f}\sin(x\Gamma t)
\label{decratbt1}\;,\\
\frac{d\Gamma[\Bzb_{\rm phys}(t)\to f]/dt}{e^{-\Gamma t}{\cal N}_f}&=&
\left(|(p/q)A_f|^2+|\overline{A}_f|^2\right)\cosh(y\Gamma t)
  -\left(|(p/q)A_f|^2-|\overline{A}_f|^2\right)\cos(x\Gamma t)\no\\
&+&2\,\re{(p/q)A_f\overline{A}^\ast_f}\sinh(y\Gamma t)
-2\,\im{(p/q)A_f\overline{A}^\ast_f}\sin(x\Gamma t)
\label{decratbt2}\; ,
\eeqa
where ${\cal N}_f$ is a common normalization factor. Decay rates to
the CP-conjugate final state $\fb$ are obtained analogously, with
${\cal N}_f = {\cal N}_{\fb}$ and the substitutions $A_f\to A_{\fb}$
and $\overline{A}_f\to\overline{A}_{\fb}$ in
Eqs.~(\ref{decratbt1},\ref{decratbt2}). Terms proportional
to $|A_f|^2$ or $|\overline{A}_f|^2 $ are associated with decays that
occur without any net $B\leftrightarrow\Bbar$ oscillation, while terms
proportional to $|(q/p)\overline{A}_f|^2$ or $|(p/q)A_f|^2$ are
associated with decays following a net oscillation. The $\sinh(y\Gamma
t)$ and $\sin(x\Gamma t)$ terms of
Eqs.~(\ref{decratbt1},\ref{decratbt2}) are associated with the
interference between these two cases. Note that, in multi-body decays, 
amplitudes are functions of phase-space variables. Interference may
be present in some regions but not others, and is strongly influenced
by resonant substructure.

One possible manifestation of CP-violating effects in meson decays
\cite{Nir:1992uv} is in the interference between a decay without
mixing, $\Bz\to f$, and a decay with mixing, $\Bz\to \Bzb\to f$ (such
an effect occurs only in decays to final states that are common to
$\Bz$ and $\Bzb$, including all CP eigenstates). It is defined by
\beq\label{cpvint}
\im{\lambda_f}\ne 0 \; ,
\eeq
with
\beq\label{deflam}
\lambda_f \equiv \frac{q}{p}\frac{\overline{A}_f}{A_f} \; .
\eeq
This form of CP violation can be observed, for example, using the
asymmetry of neutral meson decays into final CP eigenstates $f_{\CP}$
\beq\label{asyfcp}
{\cal A}_{f_{\CP}}(t)\equiv\frac{d\Gamma/dt[\Bzb_{\rm phys}(t)\to f_{\CP}]-
d\Gamma/dt[\Bz_{\rm phys}(t)\to f_{\CP}]}
{d\Gamma/dt[\Bzb_{\rm phys}(t)\to f_{\CP}]+d\Gamma/dt[\Bz_{\rm phys}(t)\to
  f_{\CP}]}\; .
\eeq
For $\Delta\Gamma = 0$ and $|q/p|=1$ (which is a good approximation
for $B$ mesons), ${\cal A}_{f_{\CP}}$ has a particularly simple form
\cite{Dunietz:1986vi,Blinov:ru,Bigi:1986vr}:
\beqa\label{asyfcpb}
{\cal A}_{f}(t)&=&S_f\sin(\Delta mt)-C_f\cos(\Delta mt),\no\\
S_f&\equiv&\frac{2\,\im{\lambda_{f}}}{1+|\lambda_{f}|^2},\ \ \ 
C_f\equiv\frac{1-|\lambda_{f}|^2}{1+|\lambda_{f}|^2}
\; ,
\eeqa

Consider the $B\to f$ decay amplitude $A_f$, and the CP conjugate
process, $\Bbar\to\fb$, with decay amplitude $\overline{A}_{\fb}$. There
are two types of phases that may appear in these decay amplitudes.
Complex parameters in any Lagrangian term that contributes to the
amplitude will appear in complex conjugate form in the CP-conjugate
amplitude. Thus their phases appear in $A_f$ and
$\overline{A}_{\overline{f}}$ with opposite signs. In the Standard
Model, these phases occur only in the couplings of the $W^\pm$ bosons
and hence are often called ``weak phases''. The weak phase of any
single term is convention dependent. However, the difference between
the weak phases in two different terms in $A_f$ is convention
independent. A second type of phase can appear in scattering or decay
amplitudes even when the Lagrangian is real. Their origin is the
possible contribution from intermediate on-shell states in the decay
process. Since these phases are generated by CP-invariant
interactions, they are the same in $A_f$ and
$\overline{A}_{\overline{f}}$. Usually the dominant rescattering is
due to strong interactions and hence the designation ``strong phases''
for the phase shifts so induced. Again, only the relative strong
phases between different terms in the amplitude are physically
meaningful.

The `weak' and `strong' phases discussed here appear in addition to
the `spurious' CP-transformation phases of Eq.~(\ref{spupha}). Those
spurious phases are due to an arbitrary choice of phase
convention, and do not originate from any dynamics or induce any \CP
violation. For simplicity, we set them to zero from here on.

It is useful to write each contribution $a_i$ to $A_f$ in three parts:
its magnitude $|a_i|$, its weak phase $\phi_i$, and its strong
phase $\delta_i$. If, for example, there are two such
contributions, $A_f = a_1 + a_2$, we have
\beqa\label{weastr}
A_f&=& |a_1|e^{i(\delta_1+\phi_1)}+|a_2|e^{i(\delta_2+\phi_2)},\no\\
\overline{A}_{\overline{f}}&=&
|a_1|e^{i(\delta_1-\phi_1)}+|a_2|e^{i(\delta_2-\phi_2)}.
\eeqa
Similarly, for neutral meson decays, it is useful to write
\beq\label{defmgam}
\Meff_{12} = |\Meff_{12}| e^{i\phi_M} \quad , \quad
\Geff_{12} = |\Geff_{12}| e^{i\phi_\Gamma} \; .
\eeq
Each of the phases appearing in Eqs.~(\ref{weastr},\ref{defmgam}) is
convention dependent, but combinations such as $\delta_1-\delta_2$,
$\phi_1-\phi_2$, $\phi_M-\phi_\Gamma$ and $\phi_M+\phi_1-\overline{\phi}_1$
(where $\overline{\phi}_1$ is a weak phase contributing to $\overline{A}_f$)
are physical. 

In the approximations that only a single weak phase contributes to decay,
$A_f=|a_f|e^{i(\delta_f+\phi_f)}$, and that
$|\Geff_{12}/\Meff_{12}|=0$, we obtain $|\lambda_f|=1$ and
the \CP asymmetries in decays to a final CP
eigenstate $f$ [Eq. (\ref{asyfcp})] with eigenvalue $\eta_f= \pm 1$
are given by
\beq\label{afcth}
{\cal A}_{f_{\CP}}(t) = \im{\lambda_f}\; \sin(\Delta m t) \; \ 
\mathrm{with}\ \   
\im{\lambda_f}=\eta_f\sin(\phi_M+2\phi_f).
\eeq
Note that the phase so measured is purely a weak phase, and no
hadronic parameters are involved in the extraction of its value from
$\im{\lambda_f}$.

%%%%%%%%%%%%%%%%%%%%%%
%%%%%%%%%%%%%%%%%%%%%%
\section{Neutrino flavor transitions}
\label{sec:nufl}
\subsection{Neutrinos in vacuum}
\label{sec:vac}
Neutrino oscillations in vacuum \cite{Pontecorvo:1957cp} arise since
neutrinos are massive and mix. In other words, the neutrino state that
is produced by electroweak interactions is not a mass eigenstate.
The weak eigenstates $\nu_\alpha$ ($\alpha=e,\mu,\tau$ denotes the
charged lepton mass eigenstates and their neutrino doublet-partners)
are linear combinations of the mass eigenstates $\nu_i$ ($i=1,2,3$):
\beq
|\nu_\alpha\rangle=U_{\alpha i}^*|\nu_i\rangle.
\eeq
After traveling a distance $L$ (or, equivalently for relativistic
neutrinos, time $t$), a neutrino originally produced with a flavor
$\alpha$ evolves as follows:
\beq
|\nu_\alpha(t)\rangle=U_{\alpha i}^*|\nu_i(t)\rangle.
\eeq
It can be detected in the charged-current interaction
$\nu_\alpha(t)N^\prime\to\ell_\beta N$ with a probability
\beq
P_{\alpha\beta}=|\langle\nu_\beta|\nu_\alpha(t)\rangle|^2=
\left|\sum_{i=1}^3\sum_{j=1}^3U_{\alpha i}^*U_{\beta
    j}\langle\nu_j(0)|\nu_i(t)\rangle\right|^2.
\eeq
We follow the analysis of ref. \cite{Gonzalez-Garcia:2002dz}. 
We use the standard approximation that $|\nu\rangle$ is a plane wave
(for a pedagogical discussion of the possible quantum mechanical
problems in this naive description of neutrino oscillations we refer
the reader to \cite{Lipkin:1999nb,Kim:1994dy}),
$|\nu_i(t)\rangle=e^{-iE_it}|\nu_i(0)\rangle$. In all cases of
interest to us, the neutrinos are relativistic:
\beq
E_i=\sqrt{p_i^2+m_i^2}\simeq p_i+\frac{m_i^2}{2E_i},
\eeq
where $E_i$ and $m_i$ are, respectively, the energy and the mass of
the neutrino mass eigenstate. Furthermore, we can assume that
$p_i\simeq p_j\equiv p\simeq E$. Then, we obtain the following
transition probability:
\beq\label{palbe}
P_{\alpha\beta}=\delta_{\alpha\beta}-4\sum_{i=1}^2\sum_{j=i+1}^3{\cal
  R}e \left(U_{\alpha i}U_{\beta i}^*U_{\alpha
    j}^*U_{\beta j}\right)\sin^2 x_{ij},
\eeq
where $x_{ij}\equiv\Delta m^2_{ij}L/(4E)$, $\Delta
m^2_{ij}=m_i^2-m_j^2$ and $L=t$ is the distance between the source
(that is, the production point of $\nu_\alpha$) and the detector (that
is, the detection point of $\nu_\beta$). In deriving Eq. (\ref{palbe})
we used the orthogonality relation $\langle\nu_j(0)|\nu_i(0)\rangle
=\delta_{ij}$. It is convenient to use the following units:
\beq
x_{ij}=1.27\ \frac{\Delta m^2_{ij}}{eV^2}\ \frac{L/E}{m/MeV}.
\eeq
The transition probability [Eq. (\ref{palbe})] has an oscillatory
behavior, with oscillation lengths
\beq
L_{0,ij}^{\rm osc}=\frac{4\pi E}{\Delta m^2_{ij}}
\eeq
and amplitude that is proportional to elements of the mixing
matrix. Thus, in order to have oscillations, neutrinos must have
different masses ($\Delta m^2_{ij}\neq0$) and they must mix
($U_{\alpha i}U_{\beta i}\neq 0$).

An experiment is characterized by the typical neutrino energy $E$ and
by the source-detector distance $L$. In order to be sensitive to a
given value of $\Delta m^2_{ij}$, the experiment has to be set up with
$E/L\approx\Delta m^2_{ij}$ ($L\sim L_{0,ij}^{\rm osc}$). The typical
values of $L/E$ for different types of neutrino sources and
experiments are summarized in Table \ref{tab:nuexp}. 

\begin{table}[t]
\caption{Characteristic values of $L$ and $E$ for various neutrino
  sources and experiments.}
\label{tab:nuexp}
\begin{center}
\begin{tabular}{cccc} \hline\hline
\rule{0pt}{1.2em}%
%\label{tab:bqqq}
%\settabs 5 \columns
Experiment & $L$ (m) & $E$ (MeV) & $\Delta m^2$ (eV$^2$)  \cr \hline
Solar & $10^{10}$ & $1$ & $10^{-10}$ \cr
Atmospheric & $10^4-10^7$ & $10^2-10^5$ & $10^{-1}-10^{-4}$ \cr
Reactor & $10^2-10^3$ & $1$ & $10^{-2}-10^{-3}$ \cr
Kamland & $10^5$ & $1$ & $10^{-5}$ \cr
Accelerator & $10^2$ & $10^3-10^4$ & $\gsim10^{-1}$ \cr
Long-baseline Accelerator & $10^5-10^6$ & $10^4$ & $10^{-2}-10^{-3}$ \cr
\hline\hline
\end{tabular}
\end{center}
\end{table}

If $(E/L)\gg\Delta m^2_{ij}$ ($L\ll L_{0,ij}^{\rm osc}$), the
oscillation does not have time to give an appreciable effect because
$\sin^2x_{ij}\ll1$. The case of $(E/L)\ll\Delta m^2_{ij}$ ($L\gg
L_{0,ij}^{\rm osc}$) requires more careful consideration. One  must
take into account that, in general, neutrino beams are not
monochromatic. Thus, rather than measuring $P_{\alpha\beta}$, the
experiments are sensitive to the average probability
\beq
\langle P_{\alpha\beta}\rangle=\delta_{\alpha\beta}
-4\sum_{i=1}^2\sum_{j=i+1}^3{\cal 
  R}e \left(U_{\alpha i}U_{\beta i}^*U_{\alpha j}^*U_{\beta j}\right)
\langle\sin^2 x_{ij}\rangle.
\eeq
For $L\gg L_{0,ij}^{\rm osc}$, the oscillation phase goes through many
cycles before the detection and is averaged to $\langle\sin^2
x_{ij}\rangle=1/2$.

For a two neutrino case,
\beq\label{nuvactwo}
P_{\alpha\beta}=\delta_{\alpha\beta}-(2\delta_{\alpha\beta}-1)\sin^22\theta\sin^2x.
\eeq

%%%%%%%%
\subsection{Neutrinos in matter}
\label{sec:mat}
\subsubsection{The effective potential}
When neutrinos propagate in dense matter, the interactions with the
medium affect their properties. These effects are either coherent or
incoherent. For purely incoherent $\nu-p$ scattering, the
characteristic cross section is very small,
\beq\label{inccs}
\sigma\sim\frac{G_F^2s}{\pi}\sim10^{-43}\ {\rm cm}^2\left(\frac{E}{1\
    MeV}\right)^2.
\eeq
The smallness of this cross section is demonstrated by the fact that
if a beam of $10^{10}$ neutrinos with $E\sim1\ MeV$ was aimed at
Earth, only one would be deflected by the Earth's matter. It may seem
then that for neutrinos matter is irrelevant. However, one must take
into account that Eq. (\ref{inccs}) does not contain the contribution
from forward elastic coherent interactions. In coherent interactions,
the medium remains unchanged and it is possible to have interference
of scattered and unscattered neutrino waves which enhances the
effect. Coherence further allows one to decouple the evolution
equation of neutrinos from the equations of the medium. In this
approximation, the effect of the medium is described by an effective
potential which depends on the density and composition of the matter
\cite{Wolfenstein:1977ue}.

Consider, for example, the evolution of $\nu_e$ in a medium with
electrons. The effective low-energy Hamiltonian describing the
relevant neutrino interactions is given by
\beq
H_W=\frac{G_F}{\sqrt{2}}\left[\overline{\nu_e}(x)
  \gamma_\alpha(1-\gamma_5)e(x)\right]
\times \left[\overline{e}(x)\gamma_\alpha(1-\gamma_5)\nu_e(x)\right].
\eeq
The effective charged-current Hamiltonian due to the electrons in the
medium is
\beqa\label{cceff}
H_C^{(e)}&=&\frac{G_F}{\sqrt{2}}\int d^3p_ef(E_e,T)\langle\langle
e(s,p_e)|\overline{e}(x)\gamma^\alpha(1-\gamma_5)\nu_e(x)
\overline{\nu_e}(x)\gamma_\alpha(1-\gamma_5)e(x)|e(s,p_e)\rangle\rangle
\no\\
&=&\frac{G_F}{\sqrt{2}}\overline{\nu_e}(x)\gamma_\alpha(1-\gamma_5)\nu_e(x)
\int d^3p_ef(E_e,T)\langle\langle
e(s,p_e)|\overline{e}(x)\gamma^\alpha
(1-\gamma_5) e(x)|e(s,p_e)\rangle\rangle,\no
\eeqa
where $s$ is the electron spin and $p_e$ its momentum. Coherence
implies that $s,p_e$ are the same for the initial and final
electrons.

Expanding the electron fields $e(x)$ in plane waves and using
$a^\dagger_s(p_e)a_s(p_e)=N_e^{(s)}(p_e)$ (the number
operator), we obtain
\beqa\label{numope}
\langle\langle e(s,p_e)&|&\overline{e}(x)\gamma^\alpha
(1-\gamma_5) e(x)|e(s,p_e)\rangle\rangle=
N_e(p_e)\frac12\sum_s\overline{u_{(s)}}(p_e)\gamma_\alpha(1-\gamma_5)u_{(s)}(p_e)\no\\
&=&\frac{N_e(p_e)}{2}{\rm Tr}\left[\frac{m_e+\not p}{2E_e}
\gamma_\alpha(1-\gamma_5)\right]=N_e(p_e)\frac{p_e^\alpha}{E_e}.
\eeqa
Isotropy implies that $\int d^3p_e\vec p_ef(E_e,T)=0$. Thus only the
$p^0$ term contributes upon integration, with $\int
d^3p_ef(E_e,T)N_e(p_e)=N_e$ (the electron number density). We obtain:
\beq
H_C^{(e)}=\frac{G_FN_e}{\sqrt{2}}\overline{\nu_e}(x)\gamma_0(1-\gamma_5)\nu_e(x).
\eeq
The effective potential for $\nu_e$ induced by its charged-current
interactions with electrons in matter is then given by
\beq\label{efpoee}
V_C=\langle \nu_e|\int d^3x
H_C^{(e)}|\nu_e\rangle=\sqrt{2}G_FN_e.
\eeq
For $\overline{\nu_e}$ the sign of $V$ is reversed. The potential can
also be expressed in terms of the matter density $\rho$:
\beq
V_C=7.6\ \frac{N_e}{N_p+N_n}\ \frac{\rho}{10^{14}\ {\rm g/cm}^3}\ eV.
\eeq
Two examples that are relevant to observations are the following:
\begin{itemize}
  \item At the Earth's core $\rho\sim10\ {\rm g/cm}^3$ and
    $V\sim10^{-13}\ eV$;
\item At the solar core $\rho\sim100\ {\rm g/cm}^3$ and
  $V\sim10^{-12}\ eV$.
  \end{itemize}

%%%%  
\subsubsection{Evolution equation}
Consider a state that is an admixture of two neutrino species,
$|\nu_e\rangle$ and $|\nu_a\rangle$ or, equivalently,  $|\nu_1\rangle$
and $|\nu_2\rangle$:
\beqa
|\Phi(x)\rangle&=&\Phi_e(x)|\nu_e\rangle+\Phi_a(x)|\nu_a\rangle\no\\
&=&\Phi_1(x)|\nu_1\rangle+\Phi_2(x)|\nu_2\rangle.
\eeqa
The evolution of $\Phi$ in a medium is described by a system of
coupled Dirac equations:
\beqa
E\Phi_1&=&\left(\frac{\hbar}{i}\gamma_0\gamma_1\frac{\partial}{\partial x}
  +\gamma_0 m_1+V_{11}\right)\Phi_1+V_{12}\Phi_2,\no\\
E\Phi_2&=&\left(\frac{\hbar}{i}\gamma_0\gamma_1\frac{\partial}{\partial x}
  +\gamma_0 m_2+V_{22}\right)\Phi_2+V_{12}\Phi_1.
\eeqa
The $V_{ij}$ terms give the effective potential for neutrino mass
eigenstates. They can be simply derived from the effective potential
for interaction eigenstates [such as $V_{ee}$ of Eq. (\ref{efpoee})]:
\beq
V_{ij}=\langle\nu_i|\int d^3x H_{\rm int}^{\rm medium}|\nu_j\rangle=
U_{i\alpha}V_{\alpha\alpha}U_{j\alpha}^*.
\eeq
We decompose the neutrino state: $\Phi_i(x)=C_i(x)\phi_i(x)$, where
$\phi_i(x)$ is the Dirac spinor part satisfying
\beq
(\gamma_0\gamma_1\{[E-V_{ii}(x)]^2-m_i^2\}^{1/2}+\gamma_0
m_i+V_{ii})=E\phi_i(x).
\eeq
We make the following approximations:
\begin{enumerate}
\item The scale over which $V$ changes is much larger than the
  microscopic wavelength of the neutrino, $(\partial V/\partial
  x)V\ll\hbar m/E^2$.
\item Expanding to first order in $V$ implies that
  $V_{12}\gamma_0\gamma_1\phi_2\simeq\phi_1$,
  $V_{12}\gamma_0\gamma_1\phi_1\simeq\phi_2$, and
  $\{[E-V_{ii}(x)]^2-m_i^2\}^{1/2}\simeq E-V_{ii}(x)-m_i^2/2E$.
\end{enumerate}
From 1 we find that the Dirac equations take the form
\beqa
EC_1\phi_1&=&\frac{\hbar}{i}\gamma_0\gamma_1\frac{\partial
  C_1}{\partial x}\phi_1 
  +(\gamma_0 m_1+V_{11})C_1\phi_1+V_{12}C_2\phi_2,\no\\
EC_2\phi_2&=&\frac{\hbar}{i}\gamma_0\gamma_1\frac{\partial
  C_2}{\partial x}\phi_2 
+(\gamma_0 m_2+V_{22})C_2\phi_2+V_{12}C_1\phi_1.
\eeqa
Then multiplying by $\gamma_0\gamma_1$ and using the equation of
motion of $\phi$ and 2, we can drop the dependence on the spinor
$\phi$ and obtain
\beqa\label{diracc}
\frac{\hbar}{i}\frac{\partial C_1}{\partial x}&=&
\left(E-V_{11}(x)-\frac{m_1^2}{2E}\right)C_1-V_{12}C_2,\no\\
\frac{\hbar}{i}\frac{\partial C_2}{\partial x}&=&
\left(E-V_{22}(x)-\frac{m_2^2}{2E}\right)C_2-V_{12}C_1.
\eeqa
Changing notations $C_{i,\alpha}(x)\to\nu_{i,\alpha}(x)$ (and
$\hbar=1$), removing the diagonal piece that is proportional to $E$,
and rotating to the flavor basis, we can rewrite Eq. (\ref{diracc}) in
matrix form \cite{Wolfenstein:1977ue}:
\beq
-i\frac{\partial}{\partial x}\pmatrix{\nu_e\cr
  \nu_a\cr}=-\frac{1}{2E}M_w^2\pmatrix{\nu_e\cr \nu_a\cr},
\eeq
where we have defined an effective mass matrix in matter,
\beq\label{hweaknu}
M_w^2=\frac12\pmatrix{m_1^2+m_2^2+4EV_e-\Delta m^2\cos2\theta
  &\Delta m^2\sin2\theta\cr \Delta m^2\sin2\theta&
  m_1^2+m_2^2+4EV_a+\Delta m^2\cos2\theta\cr},
\eeq
with $\Delta m^2=m_2^2-m_1^2$.

We define the instantaneous mass eigenstates in matter, $\nu_i^m$, as
the eigenstates of $M_w$ for a fixed value of $x$. They are related to
the interaction eigenstates by a unitary transformation,
\beq
\pmatrix{\nu_e\cr \nu_a\cr}=U(\theta_m)
\pmatrix{\nu_1^m\cr \nu_2^m\cr}=
\pmatrix{\cos\theta_m&\sin\theta_m\cr -\sin\theta_m&\cos\theta_m\cr}
\pmatrix{\nu_1^m\cr \nu_2^m\cr}.
\eeq
The eigenvalues of $M_w$, that is, the effective masses in matter, are
given by \cite{Wolfenstein:1977ue,mism}
\beq
\mu^2_{1,2}=\frac{m_1^2+m_2^2}{2}+E(V_e+V_a)\mp\frac12\sqrt{
  (\Delta m^2\cos2\theta-A)^2+(\Delta m^2\sin2\theta)^2},
\eeq
while the mixing angle in matter is given by
\beq
\tan2\theta_m=\frac{\Delta m^2\sin2\theta}{\Delta m^2\cos2\theta-A},
\eeq
where
\beq\label{defa}
A\equiv2E(V_e-V_a).
\eeq

The instantaneous mass eigenstates $\nu_i^m$ are, in general, not
energy eigenstates: they mix in the evolution. The importance of this
effect is controlled by the relative size of $4E\dot\theta_m(t)$ with
respect to $\mu_2^2(t)-\mu_1^2(t)$. When the latter is much larger
than the first, $\nu_i^m$ behave approximately as energy eigenstates
and do not mix during the evolution. This is the adiabatic transition
approximation. The adiabaticity condition reads
\beq\label{adicon}
\mu_2^2(t)-\mu_1^2(t)\gg 2EA\Delta m^2\sin2\theta\left|\dot
  A/A\right|.
\eeq
The transition probability for the adiabatic case is given by
\beq\label{peeadi}
P_{ee}(t)=\left|\sum_i U_{ei}(\theta)U_{ei}^*(\theta_p)\exp\left(-
    \frac{i}{2E}\int_{t_0}^t\mu_i^2(t^\prime)dt^\prime\right)\right|^2,
\eeq
where $\theta_p$ is the mixing angle at the production point. For
the case of two-neutrino mixing, Eq. (\ref{peeadi}) takes the form
\beq\label{peeadtwo}
P_{ee}(t)=\cos^2\theta_p\cos^2\theta+\sin^2\theta_p\sin^2\theta
+\frac12\sin2\theta_p\sin2\theta\cos\left(\frac{\delta(t)}{2E}\right),
\eeq
where
\beq
\delta(t)=\int_{t_p}^t[\mu_2^2(t^\prime)-\mu_1^2(t^\prime)]dt^\prime.
\eeq
For $\mu_2^2(t)-\mu_1^2(t)\gg E$, the last term in
Eq. (\ref{peeadtwo}) is averaged out and the survival probability
takes the form
\beq\label{peeadifin}
P_{ee}=\frac12[1+\cos2\theta_p\cos2\theta].
\eeq

The relative importance of the MSW matter term [$A$ of Eq.
(\ref{defa})] and the kinematic vacuum oscillation term in the
Hamiltonian [the off-diagonal term in Eq. (\ref{hweaknu})] can be
parametrized by the quantity $\beta_{\rm MSW}$, which represents the
ratio of matter to vacuum effects (see, for example
\cite{Bahcall:2004mz}).  From Eq. (\ref{hweaknu}) we see that the
appropriate ratio is
\beq\label{defbeta}
\beta_{\rm MSW}=\frac{2\sqrt{2}G_F n_e E_\nu}{\Delta m^2}.
\eeq
The quantity $\beta_{\rm MSW}$ is the ratio between the oscillation length in
matter and the oscillation length in vacuum. In convenient units,
$\beta_{\rm MSW}$ can be written as
\beq\label{bequan}
\beta_{\rm MSW}=0.19\left(\frac{E_\nu}{1\ MeV}\right)
\left(\frac{\mu_e\rho}{100\ {\rm g\ cm}^{-3}}\right)
\left(\frac{8\times10^{-5}\ eV^2}{\Delta m^2}\right).
\eeq
Here $\mu_e$ is the electron mean molecular weight
($\mu_e\approx0.5(1+X)$, where $X$ is the mass fraction of hydrogen)
and $\rho$ is the total density. If $\beta_{\rm MSW}\lsim\cos2\theta$,
the survival probability corresponds to vacuum averaged oscillations
[see Eq. (\ref{nuvactwo})],
\beq
P_{ee}=\left(1-\frac12 \sin^22\theta\right)\ \ \ (\beta_{\rm MSW}<\cos2\theta,\
{\rm vacuum}).
\eeq
If $\beta_{\rm MSW}>1$, the survival probability corresponds to matter
dominated oscillations [see Eq. (\ref{peeadifin})],
\beq
P_{ee}=\sin^2\theta\ \ \ (\beta_{\rm MSW}>1,\ {\rm MSW}).
\eeq
The survival probability is approximately constant in either of the
two limiting regimes, $\beta_{\rm MSW}<\cos2\theta$ and $\beta_{\rm
  MSW}>1$. There is a strong energy dependence only in the transition
region between the limiting regimes.

For the Sun, $N_e(R)=N_e(0)\exp(-R/r_0)$, with $r_0\equiv
R_\odot/10.54=6.6\times10^7\ {\rm m}=3.3\times10^{14}\ eV^{-1}$.
Then, the adiabaticity condition for the Sun reads
\beq
\frac{(\Delta
  m^2/eV^2)\sin^22\theta}{(E/MeV)\cos2\theta}\gg3\times10^{-9}.
\eeq

%%%%%%%%%%%%%%%%%%%%%
%%%%%%%%%%%%%%%%%%%%%
\tighten

%%%%%%%%%
%%%%%%%%%
\end{document}